\documentclass[english,11pt]{article}
\usepackage[T1]{fontenc}
\usepackage[utf8]{inputenc}
\usepackage[english]{babel}
\usepackage[pdftex]{graphicx}
\usepackage[width=\textwidth]{caption}
\usepackage{amsmath,amsopn,amssymb,bm,dsfont,mathrsfs,amsfonts,latexsym,amsbsy}
\usepackage{a4wide}
\usepackage{array,float, placeins,flafter, longtable,booktabs,pdflscape}
\usepackage{url}
\usepackage{color}
\usepackage{natbib}
\usepackage{lmodern}

\newcommand{\eps}{\varepsilon}
\newcommand{\DID}{\text{DID}}
\newcommand{\DIDM}{\DID_{\text{M}}}
\newcommand{\st}[1]{\texttt{#1}}
\newcommand{\0}{\bm{0}}
\newcommand{\1}{\bm{1}}

\title{Two-Way Fixed Effects and Differences-in-Differences with Heterogeneous Treatment Effects: A Survey\thanks{We are very grateful to Francesco Armillei, Kirill Borusyak, Bruno Ferman, Xavier Jaravel, Roberto Ramos, Jonathan Roth, Jann Spiess, Gonzalo Vazquez-Bare, Kaspar W\"uthrich, Jaap Abbring (the editor) and two anonymous referees for their helpful comments.}}

\author{Cl\'ement de Chaisemartin\thanks{Economics Department, Sciences Po, clement.dechaisemartin@sciencespo.fr} \and Xavier D'Haultf\oe uille\thanks{CREST-ENSAE, xavier.dhaultfoeuille@ensae.fr}}

\date{}

\begin{document}

\maketitle

    \begin{abstract}

Linear regressions with period and group fixed effects are widely used to estimate policies' effects: 26 of the 100 most cited papers published by the American Economic Review from 2015 to 2019 estimate such regressions. It has recently been shown that those regressions may produce misleading estimates, if the policy's effect is heterogeneous between groups or over time, as is often the case. This survey reviews a fast-growing literature that documents this issue, and that proposes alternative estimators robust to heterogeneous effects. We use those alternative estimators to revisit \cite{wolfers2006did}.


    \end{abstract}


  \section{Introduction}
\label{sec:introduction}

A popular method to estimate the effect of a policy, or treatment, on an outcome is to compare over time groups experiencing different evolutions of their exposure to treatment. In practice, this idea is implemented by regressing $Y_{g,t}$, the outcome in group $g$ and at period $t$, on group fixed effects, period fixed effects, and $D_{g,t}$, the treatment of group $g$ at period $t$. For instance, to measure the effect of the minimum wage on employment in the US, researchers have often regressed employment in county $g$ and year $t$ on county fixed effects, year fixed effects, and the minimum wage in county $g$ and year $t$.

\medskip
Such two-way fixed effects (TWFE) regressions are probably the most-commonly used technique in economics to measure the effect of a treatment on an outcome. \cite{de2020difference} conducted a survey of the 20 papers with the most Google Scholar citations published by the American Economic Review in 2015, and of the similarly selected papers in 2016, 2017, 2018, and 2019. Of those 100 papers, 26 have estimated at least one TWFE regression to estimate the effect of a treatment on an outcome. TWFE regressions are also very commonly used in political science, sociology, and environmental sciences.

\medskip
Researchers have long thought that TWFE estimators are equivalent to differences-in-differences (DID) estimators. With two groups and two periods, a DID estimator compares the outcome evolution from period $1$ to $2$ between a treatment group $s$ that switches from untreated to treated, and a control group $n$ that is untreated at both dates:
\begin{equation}\label{eq:DIDdebase}
\DID=Y_{s,2}-Y_{s,1}-\left(Y_{n,2}-Y_{n,1}\right).
\end{equation}
$\DID$ relies on a parallel trends assumption: in the absence of the treatment, both groups would have experienced the same outcome evolution. Specifically, for every $g\in \{s,n\}$ and $t\in \{1,2\}$, let $Y_{g,t}(0)$ and $Y_{g,t}(1)$ denote the potential outcomes in group $g$ at period $t$ without and with the treatment, respectively.\footnote{Implicitly, this notation rules out dynamic treatment effects, and assumes that groups' potential outcomes only depend on their current treatment, not on their past treatments. This restriction is not of essence to derive Equation \eqref{eq:DID2groups2periods} below, but it is of essence for some of the other results we cover, as noted later in the paper. We relax it in Section 3.2.} Parallel trends requires that the expected evolution of the untreated outcome be the same in both groups:
$$E\left[Y_{s,2}(0)-Y_{s,1}(0)\right]=E\left[Y_{n,2}(0)-Y_{n,1}(0)\right].$$
Under that assumption, $\DID$ is unbiased for the average treatment effect (ATE) in group $s$ at period $2$ (see, e.g., \cite{Abadie05}):
\begin{align}\label{eq:DID2groups2periods}
E\left[\DID\right]=&E\left[Y_{s,2}-Y_{s,1}-(Y_{n,2}-Y_{n,1})\right]\nonumber\\
=&E\left[Y_{s,2}(1)-Y_{s,1}(0)-(Y_{n,2}(0)-Y_{n,1}(0))\right]\nonumber\\
=&E\left[Y_{s,2}(1)-Y_{s,2}(0)\right]+E\left[Y_{s,2}(0)-Y_{s,1}(0)\right]-E\left[Y_{n,2}(0)-Y_{n,1}(0)\right]\nonumber\\
=&E\left[Y_{s,2}(1)-Y_{s,2}(0)\right],
\end{align}
where the last equality follows from the parallel trends assumption.
Parallel trends is partly testable, by comparing the outcome trends of groups $s$ and $n$, before group $s$ received the treatment. In practice, such pre-trends tests sometimes fail, but other times they indicate that the two groups were indeed on parallel paths before $s$ got treated.\footnote{Pre-trends tests come with caveats unveiled by a recent literature, see \cite{kahn2020promise}, \cite{bilinski2018nothing}, and \cite{roth2019pre}. Similarly, recent papers have proposed relaxations of the parallel trends assumption \citep[see, e.g.,][]{manski2018right,rambachan2019honest,freyaldenhoven2019pre}. Though we allude to it in Section \ref{sub:stagg}, this literature is mostly beyond the scope of this survey. See \cite{Roth2022} for a review.}

\medskip
Motivated by the fact that in the two-groups and two-periods design described above, $\DID$ is equal to the treatment coefficient in a TWFE regression, researchers have also estimated TWFE regressions in more complicated designs with many groups and periods, variation in treatment timing, treatments switching on and off, and/or non-binary treatments. Recent research has shown that in those more complicated designs, TWFE estimators are unbiased for an ATE if parallel trends holds, and if another assumption is satisfied: the treatment effect should be constant, between groups and over time. Unlike parallel trends, this assumption is unlikely to hold, even approximately, in most of the applications where TWFE regressions have been used. For instance, the effect of the minimum wage on employment is likely to differ in counties with highly educated workers, and in counties with less educated workers.

\medskip
The realization that one of the most commonly used empirical methods in social science relies on an often-implausible assumption has spurred a flurry of methodological papers diagnosing the seriousness of the issue, and proposing alternative estimators. This review aims to provide an overview of this recent literature, which has developed in such a quick and dynamic manner that some practitioners may have gotten lost in the whirlwind of new working papers.
We start by giving an overview of the papers that have identified TWFE's regressions lack of robustness to heterogeneous treatment effects, and that have proposed diagnostic tools practitioners may use to assess the seriousness of this issue. We then give an overview of the papers that have proposed alternative estimators robust to heterogeneous treatment effects. Finally, we revisit \cite{wolfers2006did}, a famous TWFE application, in light of the recent literature discussed in this survey. As a word of caution, note that this literature is very recent, so several of the papers we review are still working papers, which have not been through the peer-review process yet.

\medskip
Table \ref{tabsum} in the conclusion summarizes the heterogeneity-robust estimators available to applied researchers, depending on their research design. When available, the Stata and R commands implementing the diagnostics tools and alternative estimators discussed in this review are referenced, and the basic syntax of the Stata command is provided. We refer the reader to the commands' help files for further details on their syntax. Finally, the Stata code for our re-analysis of \cite{wolfers2006did}, where several of the estimators discussed in this survey are computed, is available at:

\noindent
\url{https://drive.google.com/file/d/156Fu73avBvvV_H64wePm7eW04V0jEG3K/view?usp=sharing}.

\section{TWFE regressions with heterogeneous treatment effects}\label{sec:TWFE_heteffects}
\setcounter{equation}{0}
\subsection{TWFE regressions may not identify a convex combination of treatment effects}\label{sub:setup}

We consider a panel of $G$ groups observed at $T$ periods, respectively indexed by the placeholders $g$ and $t$, which can refer to any group or time period. Typically, groups are geographical entities gathering many observations, but a group could also just be a single individual or firm. Let $\widehat{\beta}_{fe}$ denote the coefficient of $D_{g,t}$, the treatment in group $g$ at period $t$, in an OLS regression of $Y_{g,t}$, the outcome of group $g$ at period $t$, on group fixed effects, period fixed effects, and $D_{g,t}$:
\begin{equation}\label{eq:TWFE}
Y_{g,t}=\widehat{\alpha}_g+\widehat{\gamma}_t+\widehat{\beta}_{fe} D_{g,t}+\epsilon_{g,t},
\end{equation}
where $\epsilon_{g,t}$ denotes the regression residual.
We assume that the regression is unweighted, but it is sometimes weighted by $N_{g,t}$, the population of group $g$ at period $t$. The results discussed below also apply to this weighted regression, see \cite{dcDH2020}.\footnote{ \label{foot:disagg} The regression could also be estimated using more disaggregated outcome data. For instance, groups may be US counties, and one may estimate the regression using individual-level outcome measures, assigning group membership based on county of residence. This disaggregated regression is equivalent to the aggregated regression in \eqref{eq:TWFE}, provided $Y_{g,t}$ is defined as the average outcome of individuals in cell $(g,t)$, and the aggregated regression is weighted by the number of individuals in cell $(g,t)$. Accordingly, the results below also apply to disaggregated regressions, see \cite{dcDH2020}.}

\medskip
\cite{dcDH2020} show that under a parallel trends assumption on the potential outcome without treatment $Y_{g,t}(0)$,
\begin{equation}
E\left[\widehat{\beta}_{fe}\right]=E\left[\sum_{(g,t):D_{g,t}\ne 0} W_{g,t} TE_{g,t}\right].	
	\label{eq:decomp_FE}
\end{equation}
If the treatment is binary, $TE_{g,t}=Y_{g,t}(1)-Y_{g,t}(0)$, the ATE in group $g$ at time $t$. If the treatment is discrete or continuous, $TE_{g,t}=(Y_{g,t}(D_{g,t})-Y_{g,t}(0))/D_{g,t}$, the effect of moving the treatment from $0$ to $D_{g,t}$ scaled by $D_{g,t}$.\footnote{\cite{dcDH2020} derive Equation \eqref{eq:decomp_FE} assuming that groups' potential outcomes only depend on their current treatment, not on their past treatments. With dynamic effects, Equation \eqref{eq:decomp_FE} still holds if the treatment is binary and staggered, except that some of the $TE_{g,t}$s become effects of having been treated for more than one period.}
The $W_{g,t}$ are weights summing to 1,
that are proportional to and of the same sign as
\begin{equation}\label{eq:numweights}
D_{g,t}-D_{g,.}-D_{.,t}+D_{.,.},
\end{equation}
where
$D_{g,.}$ is the average treatment of group $g$ across periods, $D_{.,t}$ is the average treatment at period $t$ across groups, and $D_{.,.}$ is the average treatment across groups and periods.

\medskip
Equations \eqref{eq:decomp_FE} and \eqref{eq:numweights} have two important consequences. First, $W_{g,t}$ is in general not equal to one divided by the number of treated $(g,t)$ cells, so $\widehat{\beta}_{fe}$ may be biased for the average treatment effect across those cells, the ATT. A special case where $W_{g,t}$ is equal to one divided by the number of treated $(g,t)$ cells, and where  $\widehat{\beta}_{fe}$ is therefore unbiased for the ATT is when (i) the design is staggered, meaning that groups' treatment can only increase over time and can change at most once;\footnote{Together, (i) and (ii) imply that groups can only switch from untreated to treated, and may do so at different points in time. This is probably the definition of a staggered design many people have in mind. (i) extends the definition of a staggered design to non-binary treatments.} (ii) the treatment is binary; and  (iii) there is no variation in treatment timing: all treated groups start receiving the treatment at the same date.
However, conditions (i)-(iii) are seldom met in practice. $\widehat{\beta}_{fe}$ can also be unbiased for the ATT if one is ready to make more assumptions than just parallel trends. For instance, if one is also ready to assume that $D_{g,t}-D_{g,.}-D_{.,t}+D_{.,.}$ is uncorrelated with $TE_{g,t}$, the treatment effects that are up- and down-weighted by $\widehat{\beta}_{fe}$ do not systematically differ, and one can then show that $\widehat{\beta}_{fe}$ is unbiased for the ATT \citep[see Corollary 2 in][]{dcDH2020}.\footnote{A special case of this ``no-correlation'' condition is if the treatment effect is constant, i.e. $TE_{g,t}=\delta$ for all $(g,t)$. Then, it directly follows from Equation \eqref{eq:decomp_FE} that $E\left[\widehat{\beta}_{fe}\right]=\delta$. However, constant effect is most often an implausible assumption.} Unfortunately, this no-correlation condition is often implausible. To see this, note that $D_{g,t}-D_{g,.}-D_{.,t}+D_{.,.}$ is decreasing in $D_{g,.}$, meaning that $\widehat{\beta}_{fe}$ downweights the treatment effect of groups with the highest average treatment from period $1$ to $T$. However, groups with the largest and lowest average treatment may have systematically different treatment effects. Similarly, $D_{g,t}-D_{g,.}-D_{.,t}+D_{.,.}$ is decreasing in $D_{.,t}$, and the treatment effects at time periods with the highest average treatment may also systematically differ from the treatment effects at time periods where the average treatment is lower. In staggered adoption designs, $D_{.,t}$ is increasing in $t$ so the weights are decreasing in $t$. If the treatment effect is also monotonically increasing or decreasing in $t$, this no-correlation condition will fail. This no-correlation condition is partly testable, if one observes a proxy variable $P_{g,t}$ that is likely to be correlated with $TE_{g,t}$. Then, one can just test if $D_{g,t}-D_{g,.}-D_{.,t}+D_{.,.}$ is correlated with $P_{g,t}$.

\medskip
Second, and perhaps more worryingly, Equation \eqref{eq:numweights} implies that some of the weights $W_{g,t}$ may be negative. This means that in the minimum wage example, $\widehat{\beta}_{fe}$ could be estimating something like $3$ times the effect of the minimum wage on employment in Santa Clara county, minus $2$ times the effect in Wayne county. Then, if raising the minimum wage by one dollar decreases employment by 5\% in Santa Clara county and by 20\% in Wayne county, one would have $E\left[\widehat{\beta}_{fe}\right]=3\times -0.05-(2\times -0.2)=0.25$. $E\left[\widehat{\beta}_{fe}\right]$ would be positive, while the minimum wage's effect on employment is negative both in Santa Clara and in Wayne county. This example shows that $\widehat{\beta}_{fe}$ may not satisfy the ``no-sign reversal property'': 
$E\left[\widehat{\beta}_{fe}\right]$ could for instance be positive, even if the treatment effect is strictly negative in every $(g,t)$. This phenomenon can only arise when some of the weights $W_{g,t}$ are negative: when all those weights are positive, $\widehat{\beta}_{fe}$ does satisfy the no-sign reversal property. Note that despite its intuitive appeal and its popularity among applied researchers, the no-sign reversal property is not grounded in statistical decision theory, unlike other commonly-used criteria to discriminate estimators such as the mean-squared error. Still, it is connected to the economic concept of Pareto efficiency. If an estimator satisfies ``no-sign-reversal'', the estimand attached to it can only be positive if the treatment is not Pareto-dominated by the absence of treatment, meaning that not everybody is hurt by the treatment. Conversely, the estimand can only be negative if the treatment does not Pareto-dominate the absence of treatment. On the other hand, if an estimator does not satisfy ``no-sign-reversal'', the estimand attached to it could for instance be positive, even if the treatment is Pareto-dominated.

\medskip
Inasmuch as ``no-sign-reversal'' is a desirable property, it becomes interesting to understand when  $\widehat{\beta}_{fe}$ may satisfy it. Equation \eqref{eq:numweights} shows that with a binary treatment, the weights attached to $\widehat{\beta}_{fe}$ could all be positive. With a binary treatment, all the $(g,t)$s entering the summation in \eqref{eq:decomp_FE} must have $D_{g,t}=1$, so for a weight $W_{g,t}$ to be strictly negative, one must have $1+D_{.,.}< D_{g,.}+D_{.,t}.$
This cannot happen if $D_{g,.}+D_{.,t}\leq 1$ for every $(g,t)$. Accordingly, all the weights are likely to be positive when there is no group that is treated most of the time, and no time periods where most groups are treated. In staggered designs, this has led \cite{jakiela2021simple} to propose to drop the last periods of the data, those when $D_{.,t}$ is the highest, to mitigate or eliminate the negative weights.
One could also drop the always-treated groups, if there are any.

\medskip
On the other hand, Equation \eqref{eq:numweights} shows that with a non-binary treatment, it becomes more likely that some of the weights $W_{g,t}$ are negative. \cite{gentzkow2011} study the effect of the number of newspapers in county $g$ and year $t$ on turnout in presidential elections. Assume that in year $t$, county $g$ has $1$ newspaper ($D_{g,t}=1$), which is below its average number of newspapers across years, equal, say, to $2$ ($D_{g,.}=2$). At the same time, the average number of newspapers across counties in year $t$ is equal to 2 ($D_{.,t}=2$), which is above the average number of newspapers across all counties and years, equal, say, to $1$ ($D_{.,.}=1$). Then, it follows from \eqref{eq:numweights} that the weight assigned to the effect of newspapers in county $g$ and year $t$ is strictly negative. More generally, a necessary condition to have that all weights are positive is that in every period where the population's treatment is higher than its average across periods ($D_{.,t}\geq D_{.,.}$), the treatment of each treated group must also be larger than its average across periods ($D_{g,t}\geq D_{g,.}$ for all $g$s such that $D_{g,t}\ne 0$). This condition is likely to often fail.

\medskip
The \st{twowayfeweights} Stata \citep[see][]{twowayfeweightsStata} and R \citep[see][]{twowayfeweightsR} commands compute the weights $W_{g,t}$ in \eqref{eq:decomp_FE}.
The basic syntax of the Stata command is:

\medskip
\st{twowayfeweights outcome groupid timeid treatment, type(feTR)}

\medskip
A decomposition similar to \eqref{eq:decomp_FE} can be obtained for TWFE regressions with control variables, and for $\widehat{\beta}_{fd}$, the treatment's coefficient in a regression of the outcome's first difference on the treatment's first difference and period fixed effects. \cite{dcDH2020} also derive decompositions similar to \eqref{eq:decomp_FE}, for $\widehat{\beta}_{fe}$ and $\widehat{\beta}_{fd}$, under common trends and under the assumption that the treatment effect does not change over time. The weights in all those decompositions are also computed by the \st{twowayfeweights} Stata and R commands.

\medskip
\cite{dcDH2020} use the \st{twowayfeweights} Stata command to revisit \cite{gentzkow2011}. The authors regress the change in turnout in county $g$ between two elections on the change of the county's number of newspapers and state-year fixed effects. They find that $\widehat{\beta}_{fd}=0.0026$ (s.e. $=0.0009$): one more newspaper increases turnout by 0.26 percentage points. Using the \st{twowayfeweights} Stata package, \cite{dcDH2020} find that under parallel trends, $\widehat{\beta}_{fd}$ estimates a weighted sum of the effects of newspapers on turnout in 10,077 county$\times$election cells, where 5,472 effects are weighted positively while 4,605 are weighted negatively, and where negative weights sum to -1.43. Accordingly, $\widehat{\beta}_{fd}$ is far from estimating a convex combination of effects. The weights are negatively correlated with the election year: $\widehat{\beta}_{fd}$ is more likely to upweight newspapers' effects in early elections, and to downweight or weight negatively newspapers' effects in late elections. This may lead $\widehat{\beta}_{fd}$ to be biased if newspapers' effects change over time. Similar results apply to $\widehat{\beta}_{fe}$: more than half of the weights attached to that coefficient are negative, and negative weights sum to -0.53.

\medskip
The decomposition in \eqref{eq:decomp_FE} is the main result in \cite{dcDH2020}. Related results have appeared earlier in Theorems S1 and S2 of the Supplementary Material of \cite{deChaisemartin15c}. \cite{borusyak2016} consider the case with a binary and staggered treatment. In their Lemma 1 and Proposition 1, they assume that the treatment effect varies with the duration elapsed since one has started receiving the treatment but does not vary across groups and over time. Then, they show that $\widehat{\beta}_{fe}$ estimates a weighted sum of effects, that may assign negative weights to long-run treatment effects. Their Appendix C also contains another result related to that in Equation \eqref{eq:decomp_FE}.\footnote{Prior to that, \cite{chernozhukov2013average} had shown that one-way FE regressions may be biased for the average treatment effect, though unlike TWFE regressions they always estimate a convex combination of effects.}

\subsection{The origin of the problem: ``forbidden comparisons''}\label{sub:mechanism}

\subsubsection{Forbidden comparisons when the treatment is binary and the design is staggered}
\label{ssub:forb_st}

\cite{goodman2021difference} shows that when the treatment is binary and the design is staggered, meaning that groups can switch in but not out of treatment, we have
\begin{equation}\label{eq:GB}
\widehat{\beta}_{fe}=\sum_{g\ne g',t<t'}v_{g,g',t,t'}\DID_{g,g',t,t'},
\end{equation}
where $DID_{g,g',t,t'}$ is a DID comparing the outcome evolution of two groups $g$ and $g'$ from a pre period $t$ to a post period $t'$, and where $v_{g,g',t,t'}$ are non-negative weights summing to one, with $v_{g,g',t,t'}>0$ if and only if $g$ switches treatment between $t$ and $t'$ while $g'$ does not.\footnote{\cite{goodman2021difference} actually decomposes $\widehat{\beta}_{fe}$ as a weighted average of DIDs between cohorts of groups becoming treated at the same date, and between periods of time where their treatment remains constant. One can then further decompose his decomposition, as we do here.} Some of the $DID_{g,g',t,t'}$s in Equation \eqref{eq:GB} compare a group switching treatment from $t$ to $t'$ to a group untreated at both dates, while other $DID_{g,g',t,t'}$s compare a switching group to a group treated at both dates. The negative weights in \eqref{eq:decomp_FE} originate from this second type of DIDs.

\medskip
To see that, let us consider a simple example, first introduced by \cite{borusyak2016},\footnote{\cite{borusyak2016} have also coined the ``forbidden comparisons'' expression we borrow here.} with two groups and three periods. Group $e$, the early-treated group, is untreated at period 1 and treated at periods 2 and 3. Group $\ell$, the late-treated group, is untreated at periods 1 and 2 and treated at period 3. In this example, Equation \eqref{eq:GB} reduces to
\begin{equation}\label{eq:beta_example}
\widehat{\beta}_{fe}=(\DID_{e,\ell,1,2} +\DID_{\ell,e,2,3})/2,
\end{equation}
with
\begin{align*}
\DID_{e,\ell,1,2}& = Y_{e,2}-Y_{e,1}-\left(Y_{\ell,2}-Y_{\ell,1}\right),\\
\DID_{\ell,e,2,3}& = Y_{\ell,3}-Y_{\ell,2}-\left(Y_{e,3}-Y_{e,2}\right).
\end{align*}
$\DID_{e,\ell,1,2}$ compares the period-1-to-2 outcome evolution of group $e$, that switches from untreated to treated from period $1$ to $2$, to the outcome evolution of group $\ell$ that is untreated at both periods. $\DID_{e,\ell,1,2}$ is similar to the $\DID$ estimator in Equation \eqref{eq:DIDdebase}, and under parallel trends it is unbiased for the treatment effect in group $e$ at period $2$:
\begin{align}\label{eq:DIDSvsNT}
E\left[\DID_{e,\ell,1,2}\right]=E\left[TE_{e,2}\right].
\end{align}
$\DID_{\ell,e,2,3}$, on the other hand, compares the period-2-to-3 outcome evolution of group $\ell$, that switches from untreated to treated from period $2$ to $3$, to the outcome evolution of group $e$ that is treated at both dates. At both periods, $e$'s outcome is its treated potential outcome, which is equal to the sum of its untreated outcome and its treatment effect. Accordingly, $$Y_{e,3}-Y_{e,2}=Y_{e,3}(0)+TE_{e,3}-(Y_{e,2}(0)+TE_{e,2}).$$ On the other hand, group $\ell$ is only treated at period $3$, so
$$Y_{\ell,3}-Y_{\ell,2}=Y_{\ell,3}(0)+TE_{\ell,3}-Y_{\ell,2}(0).$$
Taking the expectation of the difference between the two previous equations,
\begin{align}\label{eq:DIDSvsAT}
E\left[\DID_{\ell,e,2,3}\right]=E\left[TE_{\ell,3}-TE_{e,3}+TE_{e,2}\right],
\end{align}
where $E\left[Y_{e,3}(0)-Y_{e,2}(0)\right]$ and $E\left[Y_{\ell,3}(0)-Y_{\ell,2}(0)\right]$ cancel out under the parallel trends assumption. Finally, it follows from Equations \eqref{eq:beta_example}, \eqref{eq:DIDSvsNT}, and \eqref{eq:DIDSvsAT} that
\begin{align}\label{eq:decompFEexample}
E\left[\widehat{\beta}_{fe}\right]&=E\left[1/2TE_{\ell,3}+TE_{e,2}-1/2TE_{e,3}\right].
\end{align}
In this simple example, Equation \eqref{eq:decomp_FE} reduces to \eqref{eq:decompFEexample}. The right-hand side of Equation \eqref{eq:decompFEexample} is a weighted sum of three ATEs where one ATE receives a negative weight. As the previous derivation shows, this negative weight comes from the fact $\widehat{\beta}_{fe}$ leverages $\DID_{\ell,e,2,3}$, a DID comparing a group switching from untreated to treated to a group treated at both periods.

\medskip
To make things more concrete, Figure \ref{fig:earlylate} below shows the actual and counterfactual outcome evolution, in a numerical example with three periods and an early and a late treated group. All treatment effects are positive: the actual outcomes, on the solid lines, are always above the counterfactual outcomes on the dashed lines. However, $\widehat{\beta}_{fe}$ is negative. $\widehat{\beta}_{fe}$ is the simple average of the DID comparing the early- to the late-treated group from period one to two, which is positive, and of the DID comparing the late- to the early-treated group from period two to three, which is negative, and larger in absolute value than the first DID. The reason why the second DID is negative is that the treatment effect of the early-treated group increases substantially from period two to three, so this group's outcome increases more than that of the late-treated group.

\begin{figure}[H]
    \begin{center}
    \caption{A numerical example with three periods,
    an early and a late treated group}
    \includegraphics{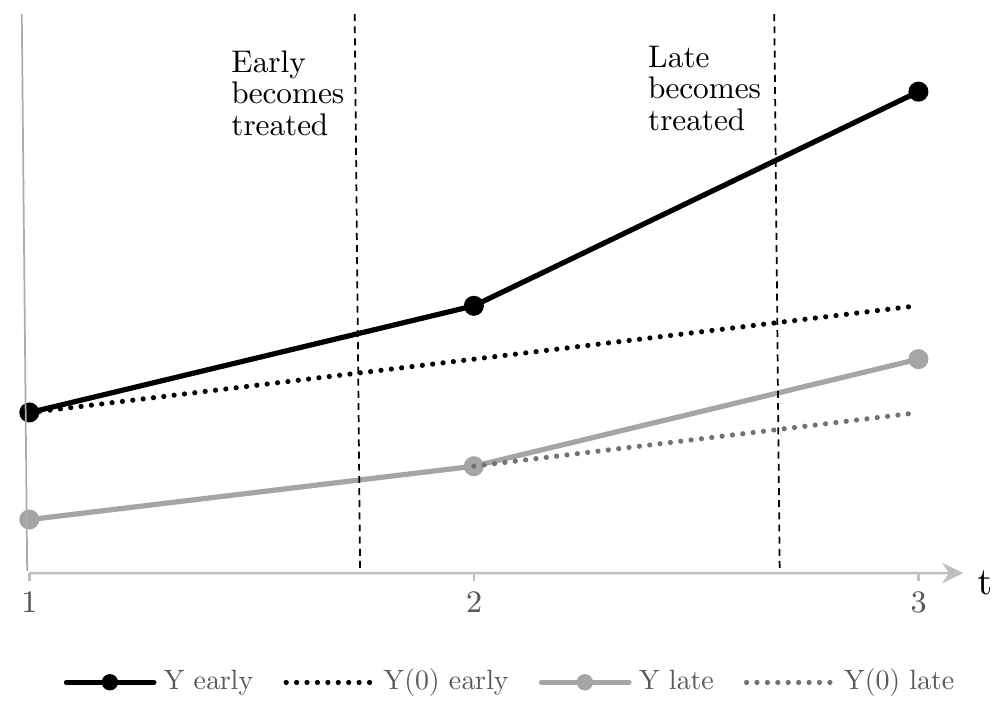}
    \label{fig:earlylate}
    \end{center}
\end{figure}

If one is ready to assume that the treatment effect does not change over time, $TE_{e,3}=TE_{e,2}$, and \eqref{eq:DIDSvsAT} simplifies to
\begin{align}\label{eq:DIDSvsAT2}
E\left[\DID_{\ell,e,2,3}\right]=E\left[TE_{\ell,3}\right].
\end{align}
Then, the negative weight in \eqref{eq:DIDSvsAT} disappears, and $\widehat{\beta}_{fe}$ estimates a weighted average of treatment effects. This extends beyond this simple example: Theorem S2 of the Web Appendix of \cite{dcDH2020} and Equation (16) of \cite{goodman2021difference} show that in staggered adoption designs with a binary treatment, $\widehat{\beta}_{fe}$ estimates a convex combination of effects, if the treatment effect does not change over time but may still vary across groups. This conclusion, however, no longer holds if the treatment is not binary or the design is not staggered. Moreover, assuming constant treatment effects over time is often implausible as this rules out both dynamic treatment effects and calendar time effects.

\medskip
The decomposition in Equation \eqref{eq:GB} is key to understand why $\widehat{\beta}_{fe}$ may not identify a convex combination of treatment effects. On the other hand, it cannot be used to assess if $\widehat{\beta}_{fe}$ does indeed estimate a convex combination of effects in a given application. Consider an example similar to that above, but with a third group $n$ that remains untreated from period $1$ to $3$. In this second example, the decomposition in \eqref{eq:GB} now indicates that $\widehat{\beta}_{fe}$ assigns a weight equal to 1/6 to DIDs comparing a switcher to a group treated at both periods. On the other hand, all the weights in \eqref{eq:decomp_FE} are positive in this second example. This phenomenon can also arise in real data sets. In the data of \cite{stevenson2006bargaining} used by \cite{goodman2021difference} in his empirical application, if one restricts the sample to states that are not always treated and to the first ten years of the panel, all the weights in \eqref{eq:decomp_FE} are positive, but the sum of the weights in \eqref{eq:GB} on DIDs comparing a switcher to a group treated at both periods is equal to $0.06$. Beyond these examples, one can show that having DIDs comparing a switcher to a group treated at both periods in \eqref{eq:GB} is necessary but not sufficient to have negative weights in \eqref{eq:decomp_FE}. Similarly, the sum of the weights on DIDs comparing a switcher to a group treated at both periods in \eqref{eq:GB} is always larger than the absolute value of the sum of the negative weights in \eqref{eq:decomp_FE}.
The reason why Equation \eqref{eq:GB} ``overestimates'' the negative weights in \eqref{eq:decomp_FE} is that as soon as there are three distinct treatment dates, there is not a unique way of decomposing $\widehat{\beta}_{fe}$ as a weighted average of DIDs, and there exists other decompositions than Equation \eqref{eq:GB} putting less weight on DIDs using a group treated at both periods as the control group.\footnote{To see that, let $t_0<t_1<t_2$ be three dates, let $e$ be an early-treated group becoming treated at $t_1$, let $\ell$ be a late-treated group becoming treated at $t_2$, and let $n$ be a group untreated yet at $t_2$. Let $\underline{v}=\min(v_{\ell,e,t_1,t_2},v_{e,n,t_0,t_2})>0$. One has
\begin{equation}\label{eq:nonuniqueDD}
\DID_{\ell,e,t_1,t_2}=\DID_{\ell,n,t_0,t_2} - \DID_{e,n,t_0,t_2} + \DID_{e,\ell,t_0,t_1}.
\end{equation}
Then, it follows from Equation \eqref{eq:nonuniqueDD} that
\begin{align}\label{eq:nonuniqueGB}
&v_{\ell,e,t_1,t_2}\DID_{\ell,e,t_1,t_2}+v_{e,n,t_0,t_2}\DID_{e,n,t_0,t_2}\nonumber\\
=&(v_{\ell,e,t_1,t_2}-\underline{v})\DID_{\ell,e,t_1,t_2}+\underline{v}\DID_{\ell,n,t_0,t_2}+\underline{v} \DID_{e,\ell,t_0,t_1}+(v_{e,n,t_0,t_2}-\underline{v})\DID_{e,n,t_0,t_2}.
\end{align}
Plugging Equation \eqref{eq:nonuniqueGB} into Equation \eqref{eq:GB} will yield a different decomposition of $\widehat{\beta}_{fe}$ as a weighted average of DIDs. But the weight on DIDs using a group treated at both periods as the control group is equal to $v_{\ell,e,t_1,t_2}$ in the left-hand-side of Equation \eqref{eq:nonuniqueGB}, and to $(v_{\ell,e,t_1,t_2}-\underline{v})$ in its right-hand side. Accordingly, this new decomposition puts strictly less weight than Equation \eqref{eq:GB} on DIDs using a group treated at both periods as the control group.}

\medskip
The \st{bacondecomp} Stata \citep[see][]{bacondecompStata} and R \citep[see][]{bacondecompR} commands compute the  $\DID_{g,g',t,t'}$s entering in \eqref{eq:GB}, the weights assigned to them, as well as the sum of the weights on $\DID_{g,g',t,t'}$s using a group treated at both periods as the control group. The basic syntax of the \st{bacondecomp} Stata command is:

\medskip
\st{bacondecomp outcome treatment, ddetail}

\subsubsection{``Forbidden comparisons'' when the design is not staggered or treatment is not binary}
\label{ssub:for_nbin}

When the treatment is not staggered or when it is not binary, $\widehat{\beta}_{fe}$ may leverage another type of comparison: it may compare the outcome evolution of a group $m$ whose treatment increases more to the outcome evolution of a group $\ell$ whose treatment increases less. In fact, with two groups $m$ and $\ell$ and two periods, one can show that
\begin{equation}\label{eq:Wald_DID}
\widehat{\beta}_{fe}=\frac{Y_{m,2}-Y_{m,1}-\left(Y_{\ell,2}-Y_{\ell,1}\right)}{D_{m,2}-D_{m,1}-\left(D_{\ell,2}-D_{\ell,1}\right)},
\end{equation}
where the right hand side of the previous display is the Wald-DID estimator studied by \cite{deChaisemartin15b}. The Wald-DID compares the outcome evolution of groups $m$ and $\ell$, and scales that comparison by the differential evolution of $m$'s and $\ell$'s treatments. \cite{deChaisemartin15b} show that the Wald-DID may not estimate a convex combination of effects, unless the treatment effect is constant over time and is the same in groups $m$ and $\ell$. This second requirement was not present in the binary and staggered case. In that case, we have seen before that if the treatment effect is constant over time, $\widehat{\beta}_{fe}$ estimates a convex combination of effects, even if the treatment effect varies between groups.

\medskip
To see that with a non-binary or non-staggered treatment $\widehat{\beta}_{fe}$ may not estimate a convex combination of effects even if the treatment effect is constant over time, let us consider a simple example. Assume that group $m$ goes from $0$ to $2$ units of treatment from period $1$ to $2$, while group $\ell$ goes from $0$ to $1$ unit. Then, the denominator of the Wald-DID is equal to $2-0-(1-0)=1$, so
\begin{equation*}
\widehat{\beta}_{fe}=Y_{m,2}-Y_{m,1}-\left(Y_{\ell,2}-Y_{\ell,1}\right).
\end{equation*}
To simplify, let us also assume that in both groups, potential outcomes are linear in the number of treatment units, with slopes that are constant over time but may differ for groups $m$ and $\ell$:
\begin{align*}
Y_{m,t}(d)=&Y_{m,t}(0)+\delta_m d\\
Y_{\ell,t}(d)=&Y_{m,t}(0)+\delta_\ell d.
\end{align*}
Then, under parallel trends,
\begin{align*}
E\left[\widehat{\beta}_{fe}\right]=&E\left[Y_{m,2}(2)-Y_{m,1}(0)-\left(Y_{\ell,2}(1)-Y_{\ell,1}(0)\right)\right]\\
=&E\left[Y_{m,2}(0)+2\delta_m-Y_{m,1}(0)-\left(Y_{\ell,2}(0)+\delta_\ell-Y_{\ell,1}(0)\right)\right]\\
=&E\left[Y_{m,2}(0)-Y_{m,1}(0)\right]-E\left[Y_{\ell,2}(0)-Y_{\ell,1}(0)\right]+2\delta_m-\delta_\ell\\
=&2\delta_m-\delta_\ell,
\end{align*}
a weighted sum of $m$ and $\ell$'s treatment effects, where group $\ell$'s effect is weighted negatively.
Intuitively, group $\ell$ is also treated at period two, and $\widehat{\beta}_{fe}$, which uses $\ell$ as a control group, subtracts its treatment effect out.
This example also shows that $\widehat{\beta}_{fe}$ may fail to identify a convex combination of effects, even without variation in treatment timing: here, both $m$ and $\ell$ start getting treated at period 2.

\medskip
To make things more concrete, Figure \ref{fig:moreless} below shows the actual and counterfactual outcome evolution, in a numerical example with two periods, a group whose treatment increases more, from $0$ to $2$ units, and a group whose treatment increases less, from $0$ to $1$ unit. All treatment effects are positive: the actual outcomes, on the solid lines, are always above the counterfactual outcomes on the dashed lines. However, $\widehat{\beta}_{fe}$, which is equal to the DID comparing the more- and the less-treated groups from period one to two, is negative. The reason why this DID is negative is that the treatment effect, per treatment unit, of the less-treated group is more than twice larger than the treatment effect of the more-treated group. Accordingly, the outcome of the less-treated group increases more, despite the fact that this group receives a twice smaller
treatment dose in period 2.
\begin{figure}[H]
    \begin{center}
    \caption{A numerical example with two periods,
    a more- and a less-treated group}
    \includegraphics{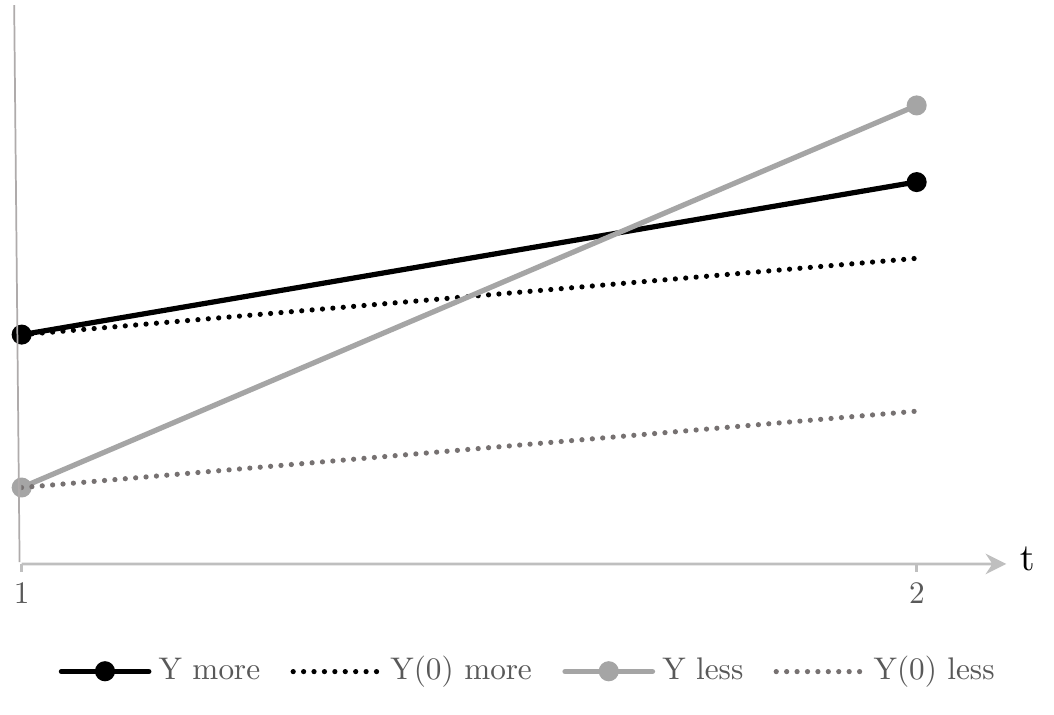}
    \label{fig:moreless}
    \end{center}
\end{figure}

\subsection{Decomposition results for other TWFE regression coefficients}\label{sub:other_TWFE}

\subsubsection{Dynamic TWFE regressions}

In staggered designs with a binary treatment, \cite{abraham2018} consider event-study regressions:
\begin{align}\label{eq:event_study}
& Y_{g,t} = \widehat{\gamma}_g + \widehat{\lambda}_t +\sum_{\ell=-K, \ell \ne -1}^L \widehat{\beta}_\ell 1\{F_g=t-\ell\}+ \eps_{g,t},
\end{align}
where $F_g$ is the first period at which group $g$ is treated. In words,
the outcome is regressed on group and period fixed effects, and relative-time indicators $1\{F_g=t-\ell\}$ equal to 1 if group $g$ started receiving the treatment $\ell$ periods ago. For $\ell\geq 0$, $\widehat{\beta}_\ell$ is supposed to estimate the cumulative effect of $\ell+1$ treatment periods. For $\ell\leq  -2$, $\widehat{\beta}_\ell$ is supposed to be a placebo coefficient testing the parallel trends assumption, by comparing the outcome trends of groups that will and will not start receiving the treatment in $|\ell|$ periods. Researchers have sometimes estimated a variant of this regression, where the first and last indicators $1\{F_g=t+K\}$ and $1\{F_g=t-L\}$ are respectively replaced by an indicator for being at least $K$ periods away from adoption ($1\{F_g\geq t+K\}$) and an indicator for having adopted at least $L$ periods ago ($1\{F_g\leq t-L\}$). Such endpoint binning is for instance recommended by \cite{Schmidheiny20}: without it, the regression implicitly assumes that the treatment no longer has any effect after $L$ periods. Instead, with endpoint binning the regression assumes that that the treatment effect is constant after $L$ periods, a more plausible assumption.

\medskip
\cite{abraham2018} show that under parallel trends, for $\ell\geq 0$,
\begin{equation}\label{eq:Abraham_Sun}
E\left[\widehat{\beta}_\ell\right]=E\left[\sum_{g}w_{g,\ell}TE_g(\ell)+\sum_{\ell'\ne\ell}\sum_{g}w_{g,\ell'}TE_g(\ell')\right],
\end{equation}
where $TE_g(\ell)$ is the cumulative effect of $\ell+1$ treatment periods in group $g$, and $w_{g,\ell}$ and $w_{g,\ell'}$ are weights such that $\sum_{g}w_{g,\ell}=1$ and $\sum_{g}w_{g,\ell'}=0$ for every $\ell'$.\footnote{Equation \eqref{eq:Abraham_Sun} follows from Proposition 3 in \cite{abraham2018}, assuming no binning and that the treatment does not have an effect after $L+1$ periods of exposure. A slight difference is that the decomposition in \cite{abraham2018} gathers groups that started receiving the treatment at the same period into cohorts. Their decomposition can then be further decomposed, as we do here.}
The first summation in the right-hand side of Equation \eqref{eq:Abraham_Sun} is a weighted sum across groups of the cumulative effect of $\ell+1$ treatment periods, with weights summing to 1 but that may be negative. This first summation resembles that in the decomposition of the ``static'' TWFE coefficient in \eqref{eq:decomp_FE}, and it implies that $\widehat{\beta}_\ell$ may be biased if the cumulative effect of $\ell+1$ treatment periods varies across groups. The second summation is a weighted sum, across $\ell'\ne \ell$ and groups, of the cumulative effect of $\ell'+1$ treatment periods in group $g$, with weights summing to 0. This second summation was not present in the decomposition of the static TWFE coefficient. Importantly, its presence implies that $\widehat{\beta}_\ell$, which is supposed to estimate the cumulative effect of $\ell+1$ treatment periods, may in fact be contaminated by the effects of $\ell'+1$ treatment periods. As $\sum_{g}w_{g,\ell'}=0$ for every $\ell'$, this second summation disappears if $TE_g(\ell')$ does not vary across groups, but it is often implausible that the treatment effect does not vary across groups.

\medskip
For $\ell\leq -2$, and without assuming parallel trends, \cite{abraham2018} show that $\widehat{\beta}_\ell$ estimates the sum of two terms. As intended, the first term measures deviations from parallel trends between groups that will and will not start receiving the treatment in $|\ell|$ periods. But the second term is similar to the second summation in the right-hand side of Equation \eqref{eq:Abraham_Sun}: a weighted sum, across $\ell'\geq 0$ and groups, of the cumulative effect of $\ell'+1$ treatment periods in group $g$, with weights summing to zero. Due to the presence of this second term, the expectation of $\widehat{\beta}_\ell$ may differ from zero even if parallel trends holds, and it may be equal to zero even if parallel trends fails. Thus, an important consequence of the results in \cite{abraham2018} is that in the presence of heterogeneous treatment effects, \eqref{eq:event_study} cannot be used to test for parallel trends.

\medskip
The \st{eventstudyweights} Stata command \citep[see][]{EVENTSTUDYWEIGHTS} computes the weights attached to event-study regressions. Its basic syntax is:

\medskip
\st{eventstudyweights  \{rel\_time\_list\}, absorb(i.groupid i.timeid) } \\
\st{cohort(first\_treatment) rel\_time(ry)},

\medskip
where \st{rel\_time\_list} is the list of relative-time indicators $1\{F_g=t-\ell\}$ included in \eqref{eq:event_study}, \st{first\_treatment} is a variable equal to the period when group $g$ got treated for the first time, and \st{ry} is a variable equal to \st{timeid} minus \st{first\_treatment}, the number of periods elapsed since group $g$ started receiving the treatment.

\medskip
Event-study regressions can only be used in staggered designs with a binary treatment. In more complicated designs where the treatment is not binary or a group's treatment can increase or decrease multiple times,
some researchers have estimated TWFE regressions of the outcome on the treatment and its first $K$ lags, the so-called distributed-lag regression. Other researchers have estimated a panel-data version of the local-projection method proposed by \cite{jorda2005estimation} for time-series data: $Y_{g,t+\ell}$ is regressed on group and period FEs and $D_{g,t}$, for $\ell \in \{0,...,K\}$. \cite{de2020difference} show that those regressions suffer from similar issues as the event-study regression: under parallel trends, the distributed-lag and local-projection regressions may produce biased estimates of the treatment's instantaneous and dynamic effects, if effects are heterogeneous across groups and over time. In particular, they do not satisfy the no-sign reversal property: one could have that the treatment's instantaneous and dynamic effects are positive in every $(g,t)$ cell, but the expectations of those regression coefficients are negative. \cite{de2020difference} also show that the panel-data version of the local-projection method may yield biased estimates even if effects are homogeneous.

\subsubsection{TWFE regressions with more than one treatment}

Another case of interest is TWFE regressions with several treatments. For instance, to estimate separately the effect of medical and recreational marijuana laws on consumption, one may regress marijuana consumption in state $g$ and year $t$ on state and year fixed effects, on whether state $g$ has a recreational marijuana law in year $t$, and on whether state $g$ has a medical law in year $t$. \cite{de2020two} show that in those regressions, the coefficient on a given treatment identifies a weighted sum of that treatment's effect across $(g,t)$s, with weights summing to $1$ but that may be negative, plus weighted sums of the effects of the other treatments in the regression, with weights summing to 0. In the example above, the coefficient on recreational laws may be contaminated by the effect of medical laws.
The weights attached to TWFE regressions with several treatments are also computed by the \st{twowayfeweights} Stata and R commands.

\section{Alternative heterogeneity-robust DID estimators}\label{sec:alternative}
\setcounter{equation}{0}

In this section, we review several recently-proposed alternatives to TWFE regressions. We restrict our attention to estimators relying on parallel trends assumptions, like TWFE regressions, but that do not restrict treatment effect heterogeneity between groups and over time, unlike TWFE regressions. This excludes papers that have assumed randomized treatment timing \citep[see, e.g., ][]{athey2021design,roth2021efficient} or sequential treatment randomization
\citep[see, e.g., ][]{bojinov2020panel}, rather than parallel trends. Intuitively, all the estimators below carefully choose valid control groups, to avoid making the ``forbidden comparisons'' that render TWFE estimators non-robust to heterogeneous treatment effects. We start by reviewing estimators ruling out dynamic effects, i.e. that assume that a group's current outcome only depends on its current treatment, before reviewing estimators that allow dynamic effects. In complicated designs, say with a continuous treatment that changes often, allowing for dynamic effects comes with a number of costs: it may result in imprecise estimators, and may complicate the interpretation of the estimated effects. Then, one may want to carefully evaluate if past treatments are indeed likely to affect the current outcome.

\subsection{Estimators ruling out dynamic effects}

With a binary treatment, \cite{dcDH2020} propose to use the $\DIDM$ estimator. With two time periods, $\DIDM$ is merely a weighted average of
\begin{align*}
\DID_+=\frac{1}{N_{0,1}}\sum_{g:D_{g,1}=0,D_{g,2}=1}(Y_{g,2}-Y_{g,1})-\frac{1}{N_{0,0}}\sum_{g:D_{g,1}=0,D_{g,2}=0}(Y_{g,2}-Y_{g,1}),
\end{align*}
and of
\begin{align*}
\DID_-=\frac{1}{N_{1,1}}\sum_{g:D_{g,1}=1,D_{g,2}=1}(Y_{g,2}-Y_{g,1})-\frac{1}{N_{1,0}}\sum_{g:D_{g,1}=1,D_{g,2}=0}(Y_{g,2}-Y_{g,1}),
\end{align*}
where for all $(d_1,d_2)\in \{0,1\}^2$, $N_{d_1,d_2}$ denotes the number of groups such that $D_{g,1}=d_1$ and $D_{g,2}=d_2$.\footnote{Implicitly, this definition of $\DID_+$ and $\DID_-$ assumes that all groups have the same sizes. The $\DIDM$ estimator can easily be extended to instances where groups have heterogeneous sizes, see \cite{dcDH2020}.}
$\DID_+$ is a DID comparing the period-one-to-two outcome evolution of groups going from untreated to treated, the ``switchers in'', and of groups untreated at both dates. It is similar to the $\DID$ estimator in Equation \eqref{eq:DIDdebase}, and it is unbiased for the treatment effect of the switching-in groups at period $2$, under a parallel trends assumption on the untreated outcome $Y_{g,t}(0)$.
$\DID_-$ is a DID comparing the period-one-to-two outcome evolution of groups treated at both dates, and of groups going from treated to untreated, the ``switchers out''. $\DID_-$ is also similar to the $\DID$ estimator in Equation \eqref{eq:DIDdebase}, switching ``treatment'' and ``non-treatment''. Then, one can show that $\DID_-$ is unbiased for the treatment effect of the switching-out groups at period $2$, under a parallel trends assumption on the treated outcome $Y_{g,t}(1)$.

\medskip
The $\DIDM$ estimator can easily be extended to applications with more than two time periods. For each pair of consecutive time periods, one can compute a $\DID_{+,t}$ estimator comparing groups going from untreated to treated from $t-1$ to $t$ to groups untreated at both dates, and a $\DID_{-,t}$ estimator comparing groups treated at $t-1$ and $t$ to groups going from treated to untreated from $t-1$ to $t$. Then, one averages the $\DID_{+,t}$ and $\DID_{-,t}$ estimators across $t$. \cite{dcDH2020} show that the resulting estimator is unbiased for the average treatment effect across all switching $(g,t)$ cells, namely cells such that $D_{g,t}\ne D_{g,t-1}$. They also propose placebo estimators to test the parallel trends assumptions underlying $\DIDM$. The placebos compare the outcome trends of switchers and non-switchers, before the switchers switch.

\medskip
With more than two time periods, the $\DIDM$ estimator may be biased if the treatment has dynamic effects. For instance, to infer the counterfactual trend that groups going from untreated to treated from $t-1$ to $t$ would have experienced without that switch, $\DID_{+,t}$ uses as controls all groups untreated at $t-1$ and $t$. However, some of those groups may have been treated, say, at $t-2$. If the treatment has dynamic effects, this past treatment may affect their period $t-1$-to-$t$ outcome evolution, thus making them potentially invalid controls. Note that if the treatment is binary and staggered, such situations cannot arise: groups untreated at $t-1$ and $t$ have been untreated all along. Accordingly, $\DIDM$ is robust to dynamic effects in binary and staggered designs.

\medskip
The $\DIDM$ estimator can easily be extended to non-binary treatments taking a finite number of values. Then, it is a weighted average, across $d$ and $t$, of DIDs comparing the $t-1$ to $t$ outcome evolution of groups whose treatment goes from $d$ to some other value from $t-1$ to $t$, and of groups with a treatment equal to $d$ at both dates, normalized by the intensity of the treatment change experienced by the switchers. For instance, in \cite{gentzkow2011}, a county going from $2$ to $4$ newspapers is compared to a county with $2$ newspapers at both dates. The multi-period DID estimator in \cite{imai2021use} is related to the $\DIDM$ estimator. It can be used with a binary treatment, to estimate the switchers-in's treatment effect.

\medskip
The $\DIDM$ estimator is computed by the \st{did\_multiplegt} Stata \citep[see][]{did_multiplegtStata} and R \citep[see][]{did_multiplegtR} commands. The basic syntax of the Stata command is:

\medskip
\st{did\_multiplegt outcome groupid timeid treatment}

\medskip
\cite{dcDH2020} compute the $\DIDM$ estimator in the \cite{gentzkow2011} example mentioned above, that studies the effect of newspapers on turnout in US presidential elections. \cite{dcDH2020} find that $\DIDM=0.0043$ (s.e. $=0.0014$), meaning that one more newspaper increases turnout by 0.43 percentage point. $\DIDM$ is 66\% larger than, and significantly different from, $\widehat{\beta}_{fd}$, the estimator reported by \cite{gentzkow2011}.

\medskip
\cite{chaisemartin2022continuous} extend the $\DIDM$ estimator to continuous treatments.  To simplify, we present their estimators in the case with two time periods, though they readily extend to the case with more periods. \cite{chaisemartin2022continuous} assume that from period one to two, the treatment of some units, hereafter referred to as the movers, changes. They also assume that the treatment of other units, hereafter referred to as the stayers, does not change. This assumption is likely to be met when the treatment is say, trade tariffs: tariffs' reforms rarely apply to all products, so it is likely that tariffs of at least some products stay constant over time. On the other hand, this assumption is unlikely to be met when the treatment is say, precipitations: geographical units never experience the exact same precipitations over two consecutive years.

\medskip
Under the assumption that there are some stayers, the estimator proposed by \cite{chaisemartin2022continuous} compares the outcome evolution of movers and stayers, with the same period-one treatment. With a continuous treatment, such comparisons can either be achieved by reweigthing stayers by propensity score weights, or by adjusting movers' outcome change using a nonparametric regression of the outcome change on the period-one treatment among the stayers. Under parallel trends assumptions, the corresponding estimands identify a weighted average of the effect, across all movers, of moving their treatment from its period-one to its period-two value, scaled by the difference between these two values. This effect is a weighted average of the slopes of movers' potential outcome function, between their period-one and period-two treatments.

\medskip
The estimators in \cite{chaisemartin2022continuous}  can be extended to the case where there are no stayers, provided there are quasi-stayers, meaning units whose treatment barely changes from period one to two. Alternatively, one could also use the estimator proposed by \cite{graham2012identification}, which compares the outcome evolution of movers and quasi stayers, but without conditioning on units' period-one treatment. Their estimator relies on a linear treatment effect assumption, unlike those in \cite{chaisemartin2022continuous}. When there are no true stayers, both estimators require choosing a bandwidth, namely the lowest treatment change below which a unit can be considered as a quasi-stayer. Neither \cite{chaisemartin2022continuous} or \cite{graham2012identification} derive an ``optimal'' bandwidth, so for now bandwidth choice is left to the discretion of the researcher. If the data has at least three periods, one could also use the correlated-random-coefficient estimator proposed by \cite{chamberlain1992efficiency}. While it allows for some treatment effect heterogeneity, that estimator relies on a linear treatment effect assumption, like the estimator in \cite{graham2012identification}.

\medskip
\cite{chaisemartin2022continuous} show that after some relabelling, some of their estimators are equivalent or nearly equivalent to estimators that had been previously proposed by \cite{deChaisemartin15b}, \cite{Abadie05}, and \cite{callaway2018}. This implies that their estimators can be computed, up to small tweaks, by the companion software for those papers. We refer the reader to \cite{chaisemartin2022continuous} for a precise description of how their estimators can be computed using existing software.

\subsection{Estimators allowing for dynamic effects when the treatment is binary and the design is staggered.}
\label{sub:stagg}

For any $t\in \{1,...,T\}$, let $\bm{0}_t$ (resp. $\bm{1}_t$) denote a vectors of $t$ zeros (resp. ones). With dynamic effects, group $g$'s outcome at time $t$ is allowed to depend on her past treatments. For any $(d_1,...,d_t)$, let $Y_{g,t}(d_1,...,d_t)$ denote group $g$'s potential outcome at period $t$ with treatments $(d_1,...,d_t)$ from period $1$ to $t$.\footnote{This notation implicitly rules out anticipation effects: the outcome cannot depend on a group's future treatment.} In particular, $Y_{g,t}(\bm{0}_t)$ is group $g$'s outcome without ever being treated from period $1$ to $t$. With dynamic effects, \cite{callaway2018} and \cite{abraham2018} have proposed to replace the parallel trends assumption on $Y_{g,t}(0)$ by a parallel trends assumption on $Y_{g,t}(\bm{0}_t)$: for all $g\ne g'$ and $t\geq 2$,
\begin{equation}\label{eq:commontrends_dynamic}
E\left[Y_{g,t}(\bm{0}_t)-Y_{g,t-1}(\bm{0}_{t-1})\right]=E\left[Y_{g',t}(\bm{0}_t)-Y_{g',t-1}(\bm{0}_{t-1})\right].
\end{equation}
We now review the estimators proposed by \cite{callaway2018}, \cite{abraham2018}, and \cite{borusyak2020revisiting} for binary and staggered treatments, under the parallel trends assumption in Equation \eqref{eq:commontrends_dynamic}.

\subsubsection{The estimators proposed by \cite{callaway2018}}
\label{ssub:CSA}

In a staggered adoption design, groups can be aggregated into cohorts that start receiving the treatment at the same period. For all $c$ and $t$, and for all $\ell \in \{0,...,t\}$ let $\overline{Y}_{c,t}$ denote the average outcome at period $t$ across groups belonging to cohort $c$, and let $\overline{Y}_{n,t}$ denote the average outcome at period $t$ across groups that remain untreated from period $1$ to $T$, hereafter referred to as the never-treated groups, assuming for now that such groups exist.
\cite{callaway2018} define their parameters of interest as
$$TE_{c,c+\ell}=E\left[\overline{Y}_{c,c+\ell}(\bm{0}_{c-1},\bm{1}_{\ell+1})-\overline{Y}_{c,c+\ell}(\bm{0}_{c+\ell})\right],$$ the average effect of having been treated for $\ell+1$ periods in the cohort that started receiving the treatment at period $c$, for every $c\in \{2,...,T\}$ and $\ell \geq 0$ such that $\ell+c\leq T$.
To estimate, say, $TE_{c,c}$, \cite{callaway2018} propose
$$\overline{\DID}_{c,0}=\overline{Y}_{c,c}-\overline{Y}_{c,c-1}-\left(\overline{Y}_{n,c}-\overline{Y}_{n,c-1}\right),$$
a DID estimator comparing the period $c-1$-to-$c$ outcome evolution in cohort $c$ and in the never-treated groups $n$.
$\overline{\DID}_{c,0}$ is unbiased for $TE_{c,c}$:
\begin{align*}
&E\left[\overline{Y}_{c,c}-\overline{Y}_{c,c-1}-\left(\overline{Y}_{n,c}-\overline{Y}_{n,c-1}\right)\right]\\
=&E\left[\overline{Y}_{c,c}(\bm{0}_{c-1},1)-\overline{Y}_{c,c-1}(\bm{0}_{c-1})-\left(\overline{Y}_{n,c}(\bm{0}_c)-\overline{Y}_{n,c-1}(\bm{0}_{c-1})\right)\right]\\
=&E\left[\overline{Y}_{c,c}(\bm{0}_{c-1},1)-\overline{Y}_{c,c}(\bm{0}_c)\right]+E\left[\overline{Y}_{c,c}(\bm{0}_{c})-\overline{Y}_{c,c-1}(\bm{0}_{c-1})-\left(\overline{Y}_{n,c}(\bm{0}_c)-\overline{Y}_{n,c-1}(\bm{0}_{c-1})\right)\right]\\
=&E\left[\overline{Y}_{c,c}(\bm{0}_{c-1},1)-\overline{Y}_{c,c}(\bm{0}_c)\right],
\end{align*}
where the last equality follows from Equation \eqref{eq:commontrends_dynamic}. More generally, to estimate $TE_{c,c+\ell}$,
\cite{callaway2018} propose
$$\overline{\DID}_{c,\ell}=\overline{Y}_{c,c+\ell}-\overline{Y}_{c,c-1}-\left(\overline{Y}_{n,c+\ell}-\overline{Y}_{n,c-1}\right),$$
a DID estimator comparing the period-$c-1$-to-$c+\ell$ outcome evolution in cohort $c$ and in the never-treated groups $n$.

\medskip
\cite{callaway2018} extend those baseline estimators in various directions. First, they propose more aggregated estimators, such as $\DID_{\ell}$, a weighted average of the $\overline{\DID}_{c,\ell}$ estimators across all cohorts reaching $\ell$ periods after their first treatment before the end of the panel. Second, they propose estimators similar to those above, but that use the not-yet-treated instead of the never-treated as controls. For instance, all groups not yet treated at period $c$ can be used as control groups in the definition of $\overline{\DID}_{c,0}$. This is very useful when there is no never-treated group: in that case, the effects $TE_{c,c+\ell}$ can still be estimated, for every $c\geq 2$ and $\ell \geq 0$ such that $\ell+c\leq U$, where $U$ is the last period when at least one group is still untreated. Even when there are never-treated groups, one may worry that such groups are less comparable to groups that get treated at some point, and researchers sometimes prefer to discard them and only leverage variation in treatment timing. Finally, even when one is fine with keeping the never-treated groups, the not-yet-treated is a larger control group, and may lead to more precise estimators. Note that in staggered adoption designs with a binary treatment, the $\DIDM$ estimator proposed by \cite{dcDH2020} also uses the not-yet-treated as controls, and is identical to the $\DID_{0}$ estimator of the instantaneous treatment effect using the not-yet-treated as controls in \cite{callaway2018}. Third, \cite{callaway2018} also propose estimators relying on a conditional parallel trends assumption.
Fourth, they suggest placebo estimators to test the parallel trends assumptions underlying their estimators. These placebos are robust to heterogeneous effects, unlike the coefficients $\widehat{\beta}_\ell$ for $\ell\leq -2$ from the event-study regression in \eqref{eq:event_study}.

\medskip
The estimators proposed by \cite{callaway2018} are computed by the \st{csdid} Stata command \citep[see][]{csdidStata}, and by the \st{did} R command \citep[see][]{didR}. The basic syntax of the Stata command is

\medskip
\st{csdid outcome, time(timeid) gvar(cohort)}

\medskip
where \st{cohort} is equal to the period when a group starts receiving the treatment.

\subsubsection{The estimators proposed by \cite{abraham2018}}
\label{ssub:SA}

\cite{abraham2018} also propose DID estimators of the cohort-and-period specific effects $TE_{c,c+\ell}$ that only rely on the parallel trends assumption in Equation \eqref{eq:commontrends_dynamic}, and that are robust to heterogeneous treatment effects. Their estimators either use the never-treated groups as controls, or the last-treated groups if there are no never-treated. With the former control group, their estimators of the $TE_{c,c+\ell}$ parameters are identical to those proposed by \cite{callaway2018} with the same control group. Operationally, they show that their estimators can be computed via a simple linear regression, which may reduce computing time. 
Unlike \cite{callaway2018}, they do not propose estimators relying on a conditional parallel trends assumption, and they also do not propose estimators using the not-yet-treated as controls.

\medskip
Their estimators are computed by the  \st{eventstudyinteract} Stata command \citep[see][]{eventstudyinteractStata}. Its basic syntax is

\medskip
\st{eventstudyinteract outcome  \{rel\_time\_list\}, absorb(i.groupid i.timeid)} \\
\st{cohort(first\_treatment)  control\_cohort(controlgroup)}

\medskip
where \st{rel\_time\_list} is the list of relative-time indicators $1\{F_g=t-\ell\}$ one would include in the event-study regression in \eqref{eq:event_study}, \st{first\_treatment} is a variable equal to the period when group $g$ got treated for the first time, and \st{controlgroup} is an indicator for the control group observations (e.g.: the never treated).

\subsubsection{The estimators proposed by \cite{borusyak2020revisiting}, \cite{gardnertwo}, and \cite{liu2021practical}}
\label{ssub:BJS}

\cite{borusyak2020revisiting}, \cite{gardnertwo}, and \cite{liu2021practical} have proposed estimators that may be more efficient than those in \cite{callaway2018} and \cite{abraham2018}, under some assumptions. We start by reviewing \cite{borusyak2020revisiting}, before discussing the connection between their results and those in \cite{gardnertwo} and \cite{liu2021practical}. The estimators in \cite{borusyak2020revisiting} can be obtained by running a TWFE regression of the outcome on group and time fixed effects, and fixed effects for every treated $(g,t)$ cell. To be concrete, if the data has $50$ groups, $10$ time periods, and $100$ treated $(g,t)$ cells, the regression has a constant and 158 fixed effects ($49$ for groups, $9$ for time periods, and $100$ for the treated $(g,t)$ cells). Under the assumptions of the Gauss-Markov theorem, the coefficients from this regression are the linear estimators of the population coefficients with the lowest variance. But under parallel trends, the population coefficient on the fixed effect for treated cell $(g,t)$ is actually equal to $TE_{g,t}$, the ATE in cell $(g,t)$, so the estimators in \cite{borusyak2020revisiting} are the linear estimators of those ATEs with the lowest variance. With estimators of $TE_{g,t}$ in hand, one can estimate $TE_{c,c+\ell}$ as the average of all the $TE_{g,t}$s such that group $g$ started receiving the treatment at period $c$ and $t=c+\ell$. Again, Gauss-Markov ensures that this estimator is the best linear estimator of $TE_{c,c+\ell}$. As the estimators in \cite{callaway2018} and \cite{abraham2018} are also linear estimators, those in \cite{borusyak2020revisiting} have a lower variance.

\medskip
A second, numerically equivalent way of computing the estimators in \cite{borusyak2020revisiting} amounts to fitting a regression of the outcome on group and time fixed effects in the sample of untreated observations, and using that regression to predict the counterfactual outcome of treated observations. Estimates of the treatment effect of those observations are then merely obtained by substracting their counterfactual to their actual outcome.
This imputation method is computationally faster than the first. It also readily generalizes to more complicated specifications, such as triple-differences, or models allowing for group-specific linear trends. Using this representation of their estimator, \cite{borusyak2020revisiting} show that it can also be used to estimate the effect of a binary and non-staggered treatment, if that treatment does not have dynamic effects. This imputation method is the one used by the \st{did\_imputation} Stata command \citep[see][]{did_imputationStata} and by the \st{didimputation} R command \citep[see][]{did_imputationR} to compute the estimators proposed by \cite{borusyak2020revisiting}. The basic syntax of the Stata command is:

\medskip
\st{did\_imputation outcome groupid timeid first\_treatment},

\medskip
where \st{first\_treatment} is a variable equal to the period when group $g$ first got treated.

\medskip
Before \cite{borusyak2020revisiting}, \cite{liu2021practical} and \cite{gardnertwo} have proposed the same imputation method as in \cite{borusyak2020revisiting},\footnote{Even before that, \cite{gobillon2016regional} have proposed a similar strategy to estimate treatment effects under a factor model.} but the result showing that the resulting estimators are efficient under the assumptions of the Gauss-Markov theorem only appears in \cite{borusyak2020revisiting}.
Note that \cite{wooldridge2021two} has also proposed an estimation strategy connected, and in some cases numerically equivalent, to that of \cite{borusyak2020revisiting}.

\subsubsection{Understanding the differences between those estimators}
\label{ssub:diffs}

Under parallel trends, the estimators in \cite{borusyak2020revisiting} may offer precision gains with respect to those in \cite{callaway2018} or \cite{abraham2018}, under the assumptions of the Gauss-Markov theorem. Those require, among other things, that the never treated potential outcomes $Y_{g,t}(\bm{0}_t)$ be independent of each other, both across groups and over time. It is, of course, often implausible that the potential outcomes of the same group are uncorrelated over time. With serial correlation, it is no longer guaranteed that the estimators in \cite{borusyak2020revisiting} will always be more efficient than those in \cite{callaway2018} and \cite{abraham2018}, but simulations in \cite{borusyak2020revisiting} suggest that one can still expect efficiency gains with moderate serial correlation.

\medskip
If trends are not exactly parallel,
the estimators in \cite{borusyak2020revisiting} may be more or less biased than those in \cite{callaway2018} or \cite{abraham2018} depending on the nature of the violation of parallel trends. \cite{borusyak2020revisiting}  do not provide a closed-form of their estimators, but one can show that with only one treated group $s$, which starts to receive the treatment at period $t_s$, their estimator of that group's treatment effect at $t_s+\ell$ is
\begin{equation}\label{eq:BJS_ex}
Y_{s,t_s+\ell}-\frac{1}{t_s-1}\sum_{k=1}^{t_s-1}Y_{s,k}-\frac{1}{G-1}\sum_{g\ne s}\left(Y_{g,t_s+\ell}-\frac{1}{t_s-1}\sum_{k=1}^{t_s-1}Y_{g,k}\right),
\end{equation}
while the estimator in \cite{callaway2018} and \cite{abraham2018} is
\begin{equation}\label{eq:CS_ex}
Y_{s,t_s+\ell}-Y_{s,t_s-1}-\frac{1}{G-1}\sum_{g\ne s}\left(Y_{g,t_s+\ell}-Y_{g,t_s-1}\right).
\end{equation}
Equation \eqref{eq:CS_ex} shows that the estimator in \cite{callaway2018} and \cite{abraham2018} use groups' $t_s-1$ outcome, the last period before $s$ gets treated, as the baseline outcome, while Equation \eqref{eq:BJS_ex} shows that the estimator in \cite{borusyak2020revisiting} instead uses the average outcome from period $1$ to $t_s-1$ as the baseline. This is why the latter estimator is often more precise.
However, it is also more biased, when parallel trends does not exactly hold and the discrepancy between groups' trends gets larger over longer horizons, as would for instance happen when there are group-specific linear trends. In such instances, \cite{roth2019pre} notes that leveraging earlier pre-treatment periods increases the bias of a DID estimator, since one makes comparisons from earlier periods. If, on the other hand, parallel trends fails due to anticipation effects arising a few periods before $t_s$, Equations \eqref{eq:BJS_ex} and \eqref{eq:CS_ex} imply that the estimator in \cite{borusyak2020revisiting} is less biased than that in \cite{callaway2018} and \cite{abraham2018}. However, these two types of violations of parallel trends may not be equally problematic. Often times, both estimators can be immunized against anticipation effects, by redefining $t_s$ as the date when the treatment was announced. On the other hand, it is often harder to immunize them against differential trends widening over time \citep[see][for further discussion]{de2020difference}.
Beyond the simple example we consider here, deriving a closed-form expression of the estimators in \cite{borusyak2020revisiting} is not straightforward. Whether the conclusions we derive in this simple example carry through to more complicated designs is thus an open question.

\medskip
If one views parallel trends as a reasonable first-order approximation rather than an assumption that holds exactly, it may make sense to investigate
how sensitive one's findings are to violations of parallel trends. To do so, one may for instance implement the partial identification approach in \cite{manski2018right} or \cite{rambachan2019honest}. The latter approach assumes that parallel trends do not hold exactly, and that the magnitude of placebo estimators is informative as to the magnitude of the bias in the actual estimators caused by differential trends. The estimators proposed by \cite{callaway2018} and \cite{abraham2018} may be more amenable to the approach in \cite{rambachan2019honest} than the estimators proposed by \cite{borusyak2020revisiting}. Consider again the same simple example as above. For any $\ell\leq t_s-2$, one can construct the following placebo estimator:
\begin{equation}\label{eq:CS_ex_placebo}
Y_{s,t_s-1}-Y_{s,t_s-\ell-2}-\frac{1}{G-1}\sum_{g\ne s}\left(Y_{g,t_s-1}-Y_{g,t_s-\ell-2}\right).
\end{equation}
This placebo compares the treated and control groups' outcome evolution, from period $t_s-\ell-2$ to $t_s-1$, namely over $\ell+1$ periods before group $s$ got treated. It exactly mimicks the estimator of group $s$'s treatment effect at period $t_s+\ell$ proposed by \cite{callaway2018} and \cite{abraham2018}, which compares the same groups, over the same number of periods. Accordingly, the magnitude of that placebo may indeed be informative as to the magnitude of the bias of the estimator in Equation \eqref{eq:CS_ex}, as requested by \cite{rambachan2019honest}. Building a placebo that would similarly mimick the estimator proposed by \cite{borusyak2020revisiting} is not feasible, precisely because that estimator leverages all pre-treatment periods to construct its baseline. See \cite{de2020difference} for more discussion of the advantages of having placebos that mimick actual estimators.

\medskip
Another difference between these approaches is that \cite{borusyak2020revisiting} impose parallel trends for every group and between every pair of consecutive time periods.\footnote{\cite{dcDH2020} and \cite{abraham2018} also impose that assumption.} \cite{callaway2018}, on the other hand, impose a weaker parallel trends assumption: from period $c$ onwards, cohort $c$ must be on the same trend as the never-treated groups, but before that cohort $c$ may have been on a different trend. The assumption in \cite{callaway2018} is the minimal assumption ensuring that all the $TE_{c,c+\ell}$ can be unbiasedly estimated, but it is conditional on the design: which groups are required to be on parallel trends at which dates depends on groups' realized treatments. It is also not testable. We refer the reader to \cite{Marcus2020} and  \cite{borusyak2020revisiting} for further discussion on the differences between parallel trends assumptions.

\medskip
Overall, whether the estimators in \cite{borusyak2020revisiting} should be preferred to those in \cite{callaway2018} and \cite{abraham2018} may depend on one's degree of confidence in the parallel trends assumption, on the type of violations of this assumption that seems more likely to arise in the application at hand, on whether it is possible to immunize the estimators against anticipation effects by redefining the treatment date as the announcement date, and on one's willingness to undertake a sensitivity analysis such as the one proposed by \cite{rambachan2019honest}. Note also that if the estimators proposed by \cite{borusyak2020revisiting}, \cite{callaway2018}, and \cite{abraham2018} are significantly different, this implies that the parallel trends assumption, at least the ``strong version'' of this assumption imposed by \cite{borusyak2020revisiting} and \cite{abraham2018}, must be violated.

\subsection{Estimators allowing for dynamic effects when the treatment is not binary or the design is not staggered.}
\label{sub:nstagg}

\cite{de2020difference} propose treatment effect estimators robust to heterogeneous and dynamic treatment effects and that can be used even if the treatment is not binary or the design is not staggered. In their survey of 26 highly cited 2015-2019 AER papers using a TWFE regression, they find that 4 have a binary treatment and a staggered design, so being able to accommodate more general designs is important. The paper's main idea is to propose a generalization of the event-study approach to such designs, by defining the event as the period where a group's treatment changes for the first time. With a binary-and-staggered treatment, the event per this definition is the period where a group gets treated, so this definition extends the standard one to general designs.

\medskip
More specifically, \cite{de2020difference} start by showing that for any group $g$ whose treatment changed for the first time at period $F_g$, the instantaneous and dynamic effects of that change can be unbiasedly estimated. Let $$\delta_{g,\ell}=E(Y_{g,F_g+\ell}-Y_{g,F_g+\ell}(D_{g,1},...,D_{g,1}))$$ be the expected difference between group $g$'s actual outcome at $F_g+\ell$ and the counterfactual ``status quo'' outcome it would have obtained if its treatment had remained equal to its period-one value from period one to $F_g+\ell$. Let $N^c_{g,\ell}$ denote the number of groups whose treatment has not changed yet at $F_g+\ell$, and with the same treatment as $g$ at period one. \cite{de2020difference} show that
\begin{equation*}
\DID_{g,\ell} =Y_{g,F_g+\ell} - Y_{g,F_g-1} -
\frac{1}{N^c_{g,\ell}}\sum_{g':D_{g',1}=D_{g,1},F_{g'} >F_g+\ell}(Y_{g',F_g+\ell} - Y_{g',F_g-1}),
\end{equation*}
a DID estimator comparing the $F_g-1$-to-$F_g+\ell$ outcome evolution between group $g$ and groups whose treatment has not changed yet at $F_g+\ell$ and with the same treatment as $g$ at period one, is unbiased for $\delta_{g,\ell}$ under parallel trends assumptions. To test those parallel trends assumptions, they propose placebo estimators comparing the outcome trends of switchers and non-switchers before the switchers switch.

\medskip
Then, \cite{de2020difference} aggregate the $\DID_{g,\ell}$ estimators into an estimator of the effect of having experienced a weakly higher amount of treatment for $\ell$ periods. For any real number $x$ and $t\in \{1,...,T\}$, let $\bm{x}_t$ denote a $1\times t$ vector with coordinates equal to $x$. When the treatment is binary, for groups untreated at period one, $D_{g,1}=0$, so
$$\delta_{g,\ell}=E(Y_{g,F_g+\ell}(\0_{F_g-1},1,D_{g,F_g+1},...,D_{g,F_g+\ell})-Y_{g,F_g+\ell}(\0_{F_g+\ell})).$$ For groups treated at period one, $D_{g,1}=1$, so $$-\delta_{g,\ell}=E(Y_{g,F_g+\ell}(\1_{F_g+\ell})-Y_{g,F_g+\ell}(\1_{F_g-1},0,D_{g,F_g+1},...,D_{g,F_g+\ell})).$$ The right-hand side of the two equations above are effects of having experienced a weakly higher amount of treatment for $\ell+1$ periods. Accordingly, the $\DID_{g,\ell}$ estimators are aggregated into a $\DID_{\ell}$ estimator, multiplying by minus one the $\DID_{g,\ell}$  of groups treated at period one. With a non-binary treatment, one can also
aggregate the $\DID_{g,\ell}$ to estimate the effect of having experienced a weakly higher amount of treatment for $\ell+1$ periods.

\medskip
Ultimately, this approach leads to an event-study graph, with the distance to the first treatment change on the $x$-axis, the $\DID_{\ell}$ estimators on the $y$-axis to the right of zero, and placebo estimators on the $y$-axis to the left of zero. This event-study graph is useful to test the parallel trends assumption, and to provide reduced-form evidence of whether weakly increasing the treatment for $\ell+1$ periods increases or decreases the outcome on average. However, interpreting the magnitude of the $\DID_{\ell}$ estimators might be complicated. For instance, with three periods and three groups such that $(D_{1,1}=0,D_{1,2}=4,D_{1,3}=1)$, $(D_{2,1}=0,D_{2,2}=2,D_{2,3}=3)$, and $(D_{2,1}=0,D_{2,2}=0,D_{2,3}=0)$, $\DID_{1}$ estimates the average of $E(Y_{1,3}(0,4,1)-Y_{1,3}(0,0,0))$ and $E(Y_{2,3}(0,2,3)-Y_{2,3}(0,0,0))$. Accordingly, $\DID_{1}$ does not estimate by how much the outcome increases on average when the treatment increases by a given amount for a given number of periods.

\medskip
To circumvent this important limitation, two strategies can be implemented. First, the reduced-form event-study graph described above can be complemented with a first-stage event-study graph, where the outcome is replaced by the treatment. The estimators on the first-stage graph show the average value of $|D_{g,F_g+\ell}-D_{g,1}|$ across all groups entering in $\DID_{\ell}$. In the example above, the first two estimates on the first-stage graph are equal to $1/2(D_{1,2}-D_{1,1}+D_{2,2}-D_{2,1})=3$ and $1/2(D_{1,3}-D_{1,1}+D_{2,3}-D_{2,1})=2$. This reflects the fact that in this example, $\DID_{1}$ is an effect produced by increasing the previous and current treatment by $3$ and $2$ units on average. Second, a weighted average across $\ell$ of the reduced-form estimators divided by a weighted average across $\ell$ of the first-stage estimators is unbiased for a parameter with a clear economic interpretation. That parameter may be used to conduct a cost-benefit analysis comparing groups' actual treatments to the status quo scenario where they would have kept all along the same treatment as in period one. In other words, that parameter can be used to determine if the policy changes that took place over the duration of the panel led to a better situation than the one that would have prevailed if no policy change had been undertaken, a natural policy question. Importantly, that parameter can also be interpreted as an average total effect per unit of treatment, where ``total effect'' refers to the sum of the instantaneous and dynamic effects of a treatment.

\medskip
The estimators proposed by \cite{de2020difference} are computed by the \st{did\_multiplegt} Stata and R commands. To compute those estimators rather than those proposed in \cite{dcDH2020}, the Stata command's basic syntax is:

\medskip
\st{did\_multiplegt outcome groupid timeid treatment, robust\_dynamic dynamic(\#)} \\
\st{average\_effect placebo(\#) longdiff\_placebo breps(\#) cluster(groupid)},

\medskip
where \st{dynamic(\#)} specifies the horizon over which effects of a first treatment switch have to be estimated, and \st{placebo(\#)} specifies the number of placebos to be estimated.

\medskip
The estimators in \cite{de2020difference} can be used with a binary treatment switching on and off, with a discrete treatment, or with a continuous and staggered treatment (groups start getting treated at different dates, with differing intensities, but once a group gets treated its treatment intensity never changes). The estimators proposed by \cite{callaway2021difference} can also accommodate continuous and staggered treatments. For continuous and non-staggered treatments, in their Section 4.3 \cite{chaisemartin2022continuous} extend their baseline estimators to allow for dynamic effects. With respect to their baseline estimators, the main difference is that when allowing for dynamic effects, fewer units can be used as controls. Without dynamic effects, at period $t$, any unit whose treatment has not changed between $t-1$ and $t$ can be used as a valid control. With dynamic effects, only units whose treatments have not changed from period 1 to $t$ can be used as valid controls. Therefore, the need for ``stayers'' becomes even stronger when allowing for dynamic effects: many units need to keep the same value of the treatment for a large number of time periods. Developing estimators robust to dynamic effects that can be used with a continuous treatment and no stayers has not been done yet and is a promising area for future research.

\medskip
The estimators in \cite{de2020difference} can, of course, also be used with a binary and staggered treatment. Without covariates in the estimation, they are then equivalent to the estimators proposed by \cite{callaway2018} using the not-yet-treated as controls. With covariates, the estimators in \cite{callaway2018} and \cite{de2020difference} differ. \cite{callaway2018} consider time-invariant covariates, and assume that trends are parallel once we condition on them. \cite{de2020difference} instead consider time-varying covariates and assume that trends are parallel once the linear effect of those time-varying covariates on the outcome is accounted for. This for instance allows them to include group-specific linear trends in the estimation. With covariates, the parallel trends conditions in \cite{callaway2018} and \cite{de2020difference} are not nested, and in principle one could combine both.

\medskip
Finally, it is worth noting that \cite{de2020two} propose estimators for the case with several treatments. They propose both estimators that generalize the $\DIDM$ estimator in \cite{dcDH2020} and rule out dynamic effects, and estimators that generalize those in \cite{callaway2018} and allow for dynamic effects.

\section{Application}
\setcounter{equation}{0}

In this section, we revisit an application with a binary and staggered treatment, thus allowing us to compute several of the heterogeneity-robust DID estimators reviewed above.
Between 1968 and 1988, 29 US states adopted a unilateral divorce law (UDL). \cite{wolfers2006did}, building upon \cite{friedberg1998did}, studies the effects of those laws on divorce rates, using a version of the event-study regression in \eqref{eq:event_study}. We use his data (\citeauthor{Wolfersdata}, \citeyear{Wolfersdata}) to revisit this question. In what follows, estimates are weighted by states' populations and standard errors are clustered at the state level, as in \cite{wolfers2006did}. As the author estimates UDLs' dynamic effects up to 15 years after adoption, in our replication we focus on heterogeneity-robust DID estimators allowing for dynamic effects, and present the estimated effects over the same horizon. We use Stata for this replication exercise, and the versions of the \st{twowayfeweights}, \st{eventstudyinteract}, \st{csdid}, \st{did\_imputation}, and \st{did\_multiplegt} commands available from the SSC repository at the end of April 2022.

\medskip
Figure \ref{fig:results} below shows the instantaneous and dynamic effects of passing a UDL, according to six estimation methods. In the top-left panel, we show the estimates from the event-study regression in \eqref{eq:event_study}, with $L=15$, $K=10$, and endpoint binning. According to this regression, UDLs increase the divorce rate on the year when the law is passed and for seven years thereafter. 11 years after those laws are passed, their effect becomes significantly negative. Those effects are consistent with those in Column (1) of Table 2 of \cite{wolfers2006did}. Our event-study regression and that in \cite{wolfers2006did} differ on two dimensions: \cite{wolfers2006did} does not include any placebo indicator for pre-adoption periods, and he includes post-adoption indicators for bins of two years (one indicator for the year when the law is passed and the year after that, one indicator for the second and third years after the law is passed, etc.). Results seem fairly robust to those specification choices. The placebo estimates are small, and individually and jointly insignificant (F-test p-value=$0.863$).

\medskip
We follow \cite{abraham2018}, and compute the weights attached to UDLs' instantaneous effect in this event-study regression.\footnote{In practice, we use the \st{twowayfeweights} Stata command, which has an option to compute the correlation between the weights and other variables that we use below.} As shown in Equation \eqref{eq:Abraham_Sun}, this coefficient can be decomposed as the sum of two terms. The first term is a weighted sum of UDL's effects in the year when they are passed, across 27 states, where all effects receive a positive weight. The weights are negatively correlated with the year variable (correlation=$-0.232$), so this first term upweights UDLs' instantaneous effects in states passing a law early, and downweights UDLs' instantaneous effects in states passing a law late. Accordingly, this first term may differ from the average instantaneous effects of UDLs if those effects vary between early- and late-adopting states, but it at least estimates a convex combination of effects. The second term is a weighted sum of UDLs' effects in the years after they are passed. 29 effects of having passed a UDL a year ago enter in that second term. 16 enter with a positive weight, and 13 enter with a negative weight. The positive and negative weights respectively sum to $0.012$ and $-0.012$. 28 effects of having passed a UDL two years ago enter in that second term. 10 effects enter with a positive weight, and 18 enter with a negative weight. The positive and negative weights respectively sum to $0.010$ and $-0.010$. Effects of having passed a UDL three, four, ..., 14, and more than 15 years ago also enter in that second term. In total, the positive and negative weights in that second term respectively sum to around $0.064$ and $-0.064$. If UDLs' dynamic effects vary across states, that second term may not be equal to zero, thus further biasing the estimated instantaneous effect in the event-study regression. However, those contamination weights are not very large, so this bias is likely to be small. Overall, this event-study regression seems fairly robust to heterogeneous treatment effects.

\medskip
In the top-centre panel of Figure \ref{fig:results}, we use the \st{eventstudyinteract} command to compute the estimators proposed by \cite{abraham2018}. The estimated effects are very similar to those in the top-left panel. This could either be due to the fact that UDLs effects are not very heterogeneous, or to the fact that the event-study regression is fairly robust to heterogeneous treatment effects, as suggested above. Interestingly, the confidence intervals are, if anything, slightly wider in the top-left than in the top-centre panel of Figure \ref{fig:results}, thus showing that heterogeneity-robust DID estimators are not always less precise than TWFE estimators. The placebos are individually insignificant. They are also substantially smaller than the estimated effects of UDLs: it does not seem that violations of parallel trends can fully account for those estimated effects.

\medskip
In the top-right panel of Figure \ref{fig:results}, we use the \st{csdid} command to compute the estimators proposed by \cite{callaway2018}, using the ``not-yet-treated'' states as the control group. The estimated effects are very similar to those in the top-centre panel. 19 states never adopt a UDL over the period under consideration, so the group of ``never-treated'' states used as controls by \st{eventstudyinteract} is quite large, and accounts for a relatively large fraction of the group of ``not-yet-treated'' states used as controls by \st{csdid}. This may explain why in this application, the two commands yield very similar estimates. Using the larger control group of ``not-yet-treated'' states also does not lead to markedly more precise estimates: the widths of the confidence intervals are similar in the two panels. The placebos produced by \st{csdid} are small and individually insignificant. The placebos are much smaller in the top-right than in the top-centre panel. This is because \st{csdid} computes first-difference placebos, comparing the outcome evolution of treated and not-yet treated states, before the treated start receiving the treatment, and between pairs of consecutive periods.\footnote{\st{csdid} has an option to compute long-difference placebos, but it returned an error when we used it.} On the other hand, \st{eventstudyinteract} computes long-difference placebos. For instance, the second placebo, shown at $t=-3$ on the graph, compares the outcome evolution of treated and never-treated states, from $F_g-1$, the period before the treated start getting treated, to $F_g-3$. See \cite{de2020difference} for a discussion of the respective advantages of long- and first-difference placebos.

\medskip
In the bottom-left panel of Figure \ref{fig:results}, we use the \st{did\_imputation} command to compute the estimators proposed  by \cite{borusyak2020revisiting}. The effects are very similar to those found by the previous two estimators. 
The confidence interval of the instantaneous effect is much tighter in the bottom-left panel than in all other panels: for that treatment effect, the estimator proposed by \cite{borusyak2020revisiting} does lead to a large precision gain. However, the opposite can hold when one considers dynamic effects. For instance, the confidence interval of the effect two years after passing a UDL is more than 50\% larger per \st{did\_imputation} than per \st{csdid}. Accordingly, the estimators proposed by \cite{borusyak2020revisiting} do not always lead to precision gains, relative to those proposed by \cite{abraham2018} or \cite{callaway2018}. The placebos produced by \st{did\_imputation} are small, individually insignificant, and jointly insignificant (F-test p-value = 0.541).\footnote{We did not report a joint test that all placebos are equal to 0 based on \st{eventstudyinteract}: this command does not readily allow to compute this test, as it does not return the covariances between the estimators. Similarly, \st{csdid} does not allow to jointly test if the placebos in Figure \ref{fig:results} are significant: it computes a joint nullity test, but for more disaggregated placebos.} Note that the placebos computed by \st{did\_imputation} are  different from those computed by the other commands. Essentially, the command estimates a TWFE regression among all the untreated $(g,t)$, with $K$ leads of the treatment. To be consistent with the other estimations, we run the command with 9 leads. Then, everything is relative to 10 periods prior to treatment, which is why the placebo estimate is set to 0 at $t=-10$ in the bottom-left panel, instead of at $t=-1$ in the other panels.

\medskip
In the bottom-centre panel of Figure \ref{fig:results}, we use the \st{did\_multiplegt} command to compute the estimators proposed by \cite{de2020difference}. The resulting estimates are extremely close to those produced by the \st{csdid} command. The only reason why the two sets of estimates are not identical is that the estimation is weighted by states' population, and the two commands seem to handle weights slightly differently. Without weighting, the two sets of estimates are identical, as expected given that there are no covariates in the estimation and we used \st{csdid} with the not-yet-treated as controls. The placebos computed by
\st{did\_multiplegt} are long-difference placebos, similar to those computed by \st{eventstudyinteract}, except that \st{did\_multiplegt} uses the not-yet-treated as controls. They are small, and individually and jointly insignificant (F-test p-value = 0.427).

\begin{figure}[H]
    \begin{center}
    \caption{Effects of Unilateral Divorce Laws, using the data in \cite{wolfers2006did}}
    \includegraphics[width=\textwidth]{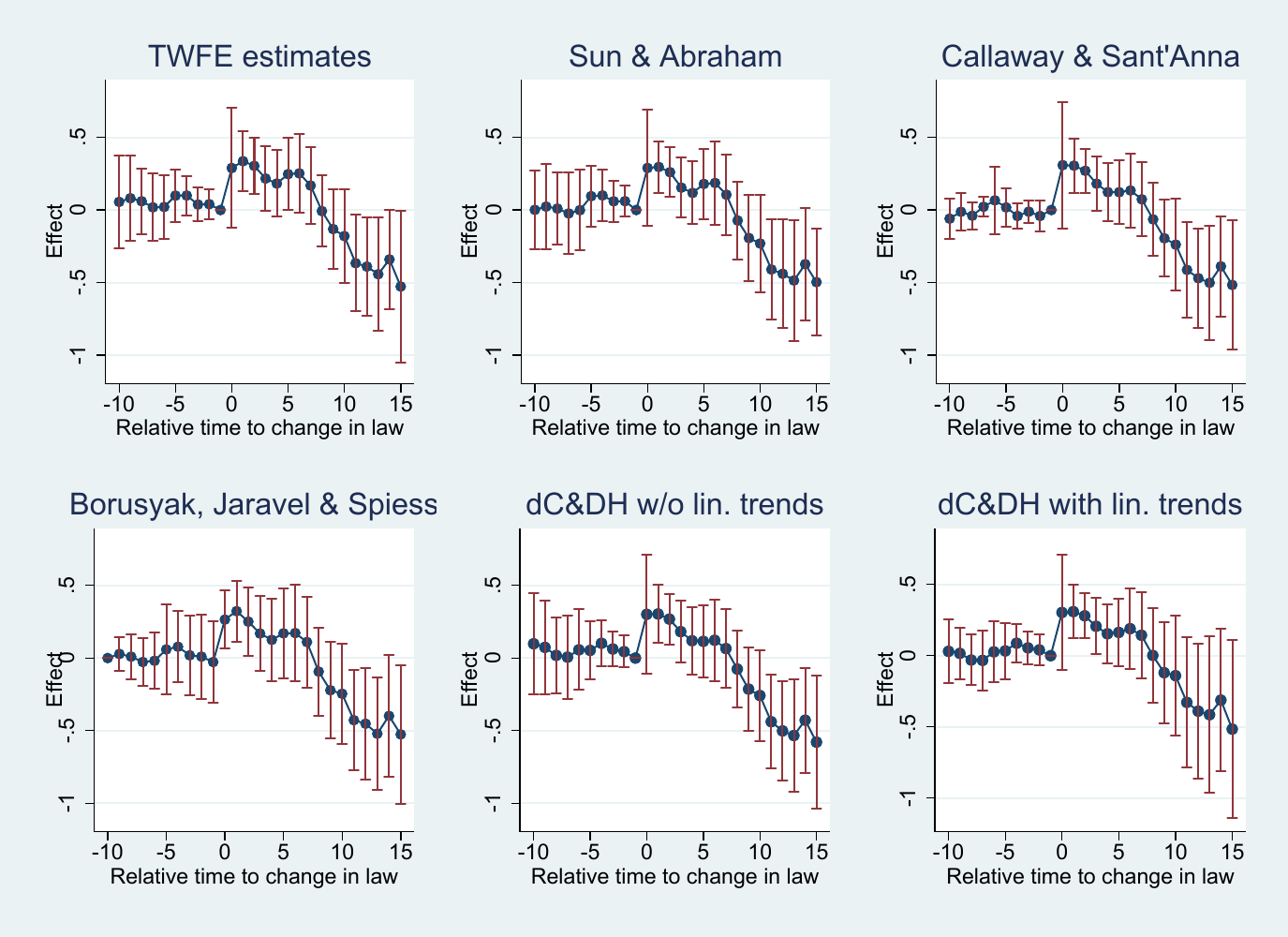}
    \label{fig:results}
    \end{center}
\footnotesize
\renewcommand{\baselineskip}{11pt}
\textbf{Note:} This figure shows the estimated effects of Unilateral Divorce Laws on the divorce rate and placebo estimates, using the data in \cite{wolfers2006did} and six estimation methods. In the top-left panel, we show estimated effects per the event-study regression in \eqref{eq:event_study}, with $L=15$, $K=10$, and endpoint binning. In the top-centre (resp. top-right, bottom-left, bottom-centre) panel, we show estimated effects per the \st{eventstudyinteract} (resp. \st{csdid}, \st{did\_imputation}, \st{did\_multiplegt}) Stata command. In the bottom-right panel, we show estimated effects per the \st{did\_multiplegt} Stata command, controlling for state-specific linear trends. All estimations are weighted by states' populations. Standard errors are clustered at the state level. 95\% confidence intervals relying on a normal approximation are shown in red.
\end{figure}

The estimates discussed so far do not control for state-specific linear trends. Whether such trends should or should not be included to estimate the effect of UDLs has been a debated issue in this literature, with \cite{friedberg1998did} arguing in their favor, and \cite{wolfers2006did} arguing that they may conflate dynamic effects. The results presented so far already suggest that including state-specific linear trends is unnecessary, as placebos are small and insignificant without them. To confirm that, we run the \st{did\_multiplegt} command again, controlling for state-specific linear trends.\footnote{\st{csdid} does not allow for group-specific trends. \st{did\_imputation} allows in principle for such trends but returned an error when such trends were added. \st{eventstudyinteract} allows for such trends.} The results, displayed in the bottom-right panel of Figure \ref{fig:results}, show that results are fairly insensitive to the inclusion of state-linear trends. If anything, adding them makes the estimated long-run effects more noisy. The only argument in favor of state-specific trends is that the placebos are slightly smaller with them, though the difference is most likely insignificant.

\medskip
Finally, to synthetize our results and obtain a point estimate that can be compared to the results in \cite{wolfers2006did}, we average UDL's effects from the year the law is passed to seven years thereafter.
The results are displayed in Table \ref{tab:results}. We do not include therein the estimates from the \st{eventstudyinteract} and \st{csdid} commands, as one cannot readily obtain the standard error of this average effect from these commands. The results show that according to all estimation methods, UDLs positively affect the divorce rate from the year the law is passed to seven years thereafter. All estimates are fairly similar to each other and point towards an increase of 20\%. The estimated standard error is substantially lower using the author's original specification, which is not surprising as it is less flexible than the other estimation methods. The estimated standard error is slightly larger using \cite{borusyak2020revisiting} than the flexible event-study regression or the estimators proposed by \cite{de2020difference}.

\begin{table}[H]
\small
\caption{\label{tab:results} The short-run effects of Unilateral Divorce Laws}
\begin{center}
\begin{tabular}{lc} \hline\hline
\\
\cite{wolfers2006did} & 0.200  \\
 & (0.056) \\
Event-study without binning pairs of years & 0.249 \\
 & (0.106) \\
\cite{borusyak2020revisiting} & 0.198 \\
& (0.129) \\
\cite{de2020difference}, no linear trends & 0.185 \\
& (0.107) \\
\cite{de2020difference}, linear trends & 0.219 \\
& (0.096) \\
\\
\hline\hline
\end{tabular}
\end{center}
\footnotesize
\renewcommand{\baselineskip}{11pt}
\textbf{Note:} This table shows the estimated effects of Unilateral Divorce Laws on the divorce rate, from 0 to 7 years after adoption, using the data in \cite{wolfers2006did}. The first set of estimates is based on the regression in Column (2) of Table 2 of \cite{wolfers2006did}. The second (resp. third, fourth) set of estimates is based on the results shown in the bottom-left (resp. bottom-centre, bottom-right) panel of Figure \ref{fig:results}. All estimations are weighted by states' populations. Standard errors, clustered at the state level, are shown beneath each estimate, between parentheses.
\end{table}

\section{Conclusion, and avenues for future research}
\setcounter{equation}{0}

The literature reviewed in this survey has shown that TWFE regressions may not always estimate a convex combination of treatment effects. In such cases, it may be hard to give them a causal interpretation, as TWFE coefficients could for instance be of a different sign than every unit's treatment effect. Table \ref{tabsum} below summarizes the alternative estimators available to applied researchers, depending on their research design and on whether they are ready or not to rule out dynamic effects. The table shows that the literature so far has mostly focused on providing alternative estimators for the case with a binary treatment and staggered adoption. Heterogeneity-robust DID estimators that can be used in more complicated designs are scarce, while many applications where TWFE regressions have been used either do not have a staggered design, or do not have a binary treatment. Developing more estimators that can be used in such designs is a promising avenue for future research. This can often be done by building upon the insights gained from studying the binary-and-staggered case. For instance, the estimators proposed by \cite{de2020difference} build upon those proposed by \cite{callaway2018} for the binary-and-staggered case. We hope that the whirlwind of DID working papers shall continue, till heterogeneity-robust DID estimators are as widely applicable as TWFE regressions.

\medskip
It is also important to stress that at this stage, it is still unclear whether researchers should systematically abandon TWFE estimators. Those estimators sometimes estimate a convex combination of effects under the parallel trends assumption, they may estimate the ATT if the weights attached to them are uncorrelated with the treatment effects $TE_{g,t}$, and they often have a lower variance than the heterogeneity-robust estimators reviewed in the previous section. While there are examples where TWFE and heterogeneity-robust DID estimators are economically and statistically different \citep[see, e.g., the empirical examples in][]{dcDH2020,de2020difference,de2020two,baker2022much}, the previous section also shows a data set where TWFE and heterogeneity-robust DID estimators lead to very similar conclusions. Understanding the circumstances where TWFE and heterogeneity-robust DID estimators are more likely to differ is an important question. We conjecture that differences are likely to be larger in complicated designs (e.g.: a non-binary treatment that can turn on and off multiple times, or several treatments) than in simple designs (e.g.: a single binary and staggered treatment). This conjecture
is based on our discussion of Equation \eqref{eq:numweights} in Section 3. This is also a pattern we found when computing TWFE and heterogeneity-robust DID estimators in four different data sets, in the empirical examples of this survey and of \citeauthor{dcDH2020} (\citeyear{dcDH2020,de2020difference,de2020two}). But those examples are not enough to draw general conclusions: a systematic comparison of TWFE and heterogeneity-robust DID estimators in a broad set of applications is in order.

\medskip
Analyzing estimators' robustness to heterogeneous treatment effects is important, as the assumption that all units are affected in the same way by a treatment is seldom credible. In this survey, we have focused on estimators relying on parallel trends assumptions, but this question is also relevant for other estimators. See for instance \cite{sloczynski2020should} and \cite{blandhol2022tsls} for instrumental variables estimators with covariates. More closely related to our set-up, the impact of heterogeneous treatment effects in the ``group fixed-effects'' model of \cite{bonhomme2015grouped} remains to be studied.

\begin{table}
\small
\caption{\label{tabsum} A summary of available heterogeneity-robust DID estimators}
\begin{center}
\begin{tabular}{llll} \hline \hline \\
\multicolumn{4}{c}{\textbf{Panel A: Estimators ruling out dynamic effects}}\\
\multicolumn{4}{c}{\textit{Can be used when outcome unaffected by past treatments}}\\
\\
\textit{Treatment} & \textit{Estimators available} & \textit{Stata commands} & \textit{See:} \\
\\
Binary   & \cite{dcDH2020} & \st{did\_multiplegt} & 3.1 \\
  & \cite{imai2021use} &  & 3.1 \\
& \cite{borusyak2020revisiting} & \st{did\_imputation} & \ref{ssub:BJS} \\
\\
Discrete  & \cite{dcDH2020} & \st{did\_multiplegt} & 3.1 \\
\\
Continuous, & & & \\
with stayers & \cite{chaisemartin2022continuous} & See Section 3.1 & 3.1\\
\\
Continuous,  & \cite{chaisemartin2022continuous} & See Section 3.1 & 3.1 \\
without stayers & \cite{graham2012identification} & \st{gmm} & 3.1 \\
& \cite{chamberlain1992efficiency} & \st{gmm} & 3.1 \\
\\
Several treatments & \cite{de2020two} & \st{did\_multiplegt} & \ref{sub:nstagg} \\
\\
\multicolumn{4}{c}{\textbf{Panel B: Estimators allowing dynamic effects}}\\
\multicolumn{4}{c}{\textit{Can be used when outcome affected by past treatments}}\\
\\
\textit{Treatment} & \textit{Estimators available} & \textit{Stata  commands} & \textit{See:} \\
\\
Binary and    & \cite{callaway2018} & \st{csdid} &  \ref{ssub:CSA}\\
staggered  & \cite{abraham2018}  & \st{eventstudyinteract} &  \ref{ssub:SA}\\
  & \cite{borusyak2020revisiting} & \st{did\_imputation} & \ref{ssub:BJS}\\
  & \cite{de2020difference} & \st{did\_multiplegt} & \ref{sub:nstagg}\\
\\
Binary or discrete, & & & \\
non-staggered  & \cite{de2020difference} & \st{did\_multiplegt} & \ref{sub:nstagg} \\
\\
Continuous and & \cite{de2020difference} & \st{did\_multiplegt} & \ref{sub:nstagg} \\
staggered  &  \cite{callaway2021difference} & & \ref{sub:nstagg} \\
\\
Continuous and & & & \\
non-staggered, & & & \\
with stayers & \cite{chaisemartin2022continuous} & See paper & \ref{sub:nstagg} \\
\\
Continuous and & & & \\
non-staggered, & & & \\
without stayers & No estimator available yet & & \\
\\
Several treatments & \cite{de2020two} & \st{did\_multiplegt} & \ref{sub:nstagg} \\
\\
\hline \hline
\end{tabular}

\end{center}

\footnotesize
\renewcommand{\baselineskip}{11pt}
\textbf{Note:} All the Stata commands have R equivalents with the same name, except \st{eventstudyinteract} that does not have an R equivalent, and \st{csdid} whose R equivalent is called \st{did}. The table's last column indicates the section of the paper where the estimator is described.
\end{table}

\newpage


\bibliographystyle{chicago}
\bibliography{biblio}

\begin{thebibliography}{}

\bibitem[\protect\citeauthoryear{Abadie}{Abadie}{2005}]{Abadie05}
Abadie, A. (2005, 01).
\newblock Semiparametric difference-in-differences estimators.
\newblock {\em Review of Economic Studies\/}~{\em 72\/}(1), 1--19.

\bibitem[\protect\citeauthoryear{Athey and Imbens}{Athey and
  Imbens}{2022}]{athey2021design}
Athey, S. and G.~W. Imbens (2022).
\newblock Design-based analysis in difference-in-differences settings with
  staggered adoption.
\newblock {\em Journal of Econometrics\/}~{\em 226}, 62--79.

\bibitem[\protect\citeauthoryear{Baker, Larcker, and Wang}{Baker
  et~al.}{2022}]{baker2022much}
Baker, A.~C., D.~F. Larcker, and C.~C. Wang (2022).
\newblock How much should we trust staggered difference-in-differences
  estimates?
\newblock {\em Journal of Financial Economics\/}~{\em 144\/}(2), 370--395.

\bibitem[\protect\citeauthoryear{Bilinski and Hatfield}{Bilinski and
  Hatfield}{2018}]{bilinski2018nothing}
Bilinski, A. and L.~A. Hatfield (2018).
\newblock Nothing to see here? non-inferiority approaches to parallel trends
  and other model assumptions.
\newblock arXiv preprint arXiv:1805.03273.

\bibitem[\protect\citeauthoryear{Blandhol, Bonney, Mogstad, and
  Torgovitsky}{Blandhol et~al.}{2022}]{blandhol2022tsls}
Blandhol, C., J.~Bonney, M.~Mogstad, and A.~Torgovitsky (2022).
\newblock When is tsls actually late?
\newblock NBER working paper 29709.

\bibitem[\protect\citeauthoryear{Bojinov, Rambachan, and Shephard}{Bojinov
  et~al.}{2021}]{bojinov2020panel}
Bojinov, I., A.~Rambachan, and N.~Shephard (2021).
\newblock Panel experiments and dynamic causal effects: A finite population
  perspective.
\newblock {\em Quantitative Economics\/}~{\em 12}, 1171–1196.

\bibitem[\protect\citeauthoryear{Bonhomme and Manresa}{Bonhomme and
  Manresa}{2015}]{bonhomme2015grouped}
Bonhomme, S. and E.~Manresa (2015).
\newblock Grouped patterns of heterogeneity in panel data.
\newblock {\em Econometrica\/}~{\em 83\/}(3), 1147--1184.

\bibitem[\protect\citeauthoryear{Borusyak}{Borusyak}{2021}]{did_imputationStata}
Borusyak, K. (2021, June).
\newblock {DID\_IMPUTATION: Stata module to perform treatment effect estimation
  and pre-trend testing in event studies}.

\bibitem[\protect\citeauthoryear{Borusyak and Jaravel}{Borusyak and
  Jaravel}{2017}]{borusyak2016}
Borusyak, K. and X.~Jaravel (2017).
\newblock Revisiting event study designs.
\newblock Working Paper.

\bibitem[\protect\citeauthoryear{Borusyak, Jaravel, and Spiess}{Borusyak
  et~al.}{2021}]{borusyak2020revisiting}
Borusyak, K., X.~Jaravel, and J.~Spiess (2021).
\newblock Revisiting event study designs: Robust and efficient estimation.
\newblock arXiv preprint arXiv:2108.12419.

\bibitem[\protect\citeauthoryear{Butts}{Butts}{2021}]{did_imputationR}
Butts, K. (2021, August).
\newblock {didimputation: Imputation Estimator from Borusyak, Jaravel, and
  Spiess (2021) in R}.

\bibitem[\protect\citeauthoryear{Callaway, Goodman-Bacon, and
  Sant'Anna}{Callaway et~al.}{2021}]{callaway2021difference}
Callaway, B., A.~Goodman-Bacon, and P.~H. Sant'Anna (2021).
\newblock Difference-in-differences with a continuous treatment.
\newblock arXiv preprint arXiv:2107.02637.

\bibitem[\protect\citeauthoryear{Callaway and Sant'Anna}{Callaway and
  Sant'Anna}{2021}]{callaway2018}
Callaway, B. and P.~H. Sant'Anna (2021).
\newblock Difference-in-differences with multiple time periods.
\newblock {\em Journal of Econometrics\/}~{\em 225}, 200--230.

\bibitem[\protect\citeauthoryear{Chamberlain}{Chamberlain}{1992}]{chamberlain1992efficiency}
Chamberlain, G. (1992).
\newblock Efficiency bounds for semiparametric regression.
\newblock {\em Econometrica\/}~{\em 60\/}(3), 567--596.

\bibitem[\protect\citeauthoryear{Chernozhukov, Fern{\'a}ndez-Val, Hahn, and
  Newey}{Chernozhukov et~al.}{2013}]{chernozhukov2013average}
Chernozhukov, V., I.~Fern{\'a}ndez-Val, J.~Hahn, and W.~Newey (2013).
\newblock Average and quantile effects in nonseparable panel models.
\newblock {\em Econometrica\/}~{\em 81\/}(2), 535--580.

\bibitem[\protect\citeauthoryear{de~Chaisemartin and
  D'Haultf{\oe}uille}{de~Chaisemartin and
  D'Haultf{\oe}uille}{2015}]{deChaisemartin15c}
de~Chaisemartin, C. and X.~D'Haultf{\oe}uille (2015).
\newblock Fuzzy differences-in-differences.
\newblock ArXiv e-prints, eprint 1510.01757v2.

\bibitem[\protect\citeauthoryear{de~Chaisemartin and
  D'Haultf{\oe}uille}{de~Chaisemartin and
  D'Haultf{\oe}uille}{2018}]{deChaisemartin15b}
de~Chaisemartin, C. and X.~D'Haultf{\oe}uille (2018).
\newblock Fuzzy differences-in-differences.
\newblock {\em The Review of Economic Studies\/}~{\em 85\/}(2), 999--1028.

\bibitem[\protect\citeauthoryear{de~Chaisemartin and
  D'Haultf{\oe}uille}{de~Chaisemartin and D'Haultf{\oe}uille}{2020}]{dcDH2020}
de~Chaisemartin, C. and X.~D'Haultf{\oe}uille (2020).
\newblock Two-way fixed effects estimators with heterogeneous treatment
  effects.
\newblock {\em American Economic Review\/}~{\em 110\/}(9), 2964--2996.

\bibitem[\protect\citeauthoryear{de~Chaisemartin and
  D'Haultf{\oe}uille}{de~Chaisemartin and
  D'Haultf{\oe}uille}{2021a}]{de2020difference}
de~Chaisemartin, C. and X.~D'Haultf{\oe}uille (2021a).
\newblock Difference-in-differences estimators of intertemporal treatment
  effects.
\newblock arXiv preprint arXiv:2007.04267.

\bibitem[\protect\citeauthoryear{de~Chaisemartin and
  D'Haultf{\oe}uille}{de~Chaisemartin and
  D'Haultf{\oe}uille}{2021b}]{de2020two}
de~Chaisemartin, C. and X.~D'Haultf{\oe}uille (2021b).
\newblock Two-way fixed effects regressions with several treatments.
\newblock arXiv preprint arXiv:2012.10077.

\bibitem[\protect\citeauthoryear{de~Chaisemartin, D'Haultf{\oe}uille, and
  Deeb}{de~Chaisemartin et~al.}{2019}]{twowayfeweightsStata}
de~Chaisemartin, C., X.~D'Haultf{\oe}uille, and A.~Deeb (2019, February).
\newblock {twowayfeweights: Estimation of the Weights Attached to the Two-Way
  Fixed Effects Regressions in Stata}.

\bibitem[\protect\citeauthoryear{de~Chaisemartin, D'Haultf{\oe}uille, and
  Guyonvarch}{de~Chaisemartin et~al.}{2019}]{did_multiplegtStata}
de~Chaisemartin, C., X.~D'Haultf{\oe}uille, and Y.~Guyonvarch (2019, May).
\newblock {did\_multiplegt: DID Estimation with Multiple Groups and Periods in
  Stata}.

\bibitem[\protect\citeauthoryear{de~Chaisemartin, D’Haultfoeuille, Pasquier,
  and Vazquez-Bare}{de~Chaisemartin et~al.}{2022}]{chaisemartin2022continuous}
de~Chaisemartin, C., X.~D’Haultfoeuille, F.~Pasquier, and G.~Vazquez-Bare
  (2022).
\newblock Difference-in-differences estimators of the effect of a continuous
  treatment.
\newblock arXiv preprint arXiv:2201.06898.

\bibitem[\protect\citeauthoryear{Flack and Edward}{Flack and
  Edward}{2020}]{bacondecompR}
Flack, E. and Edward (2020, January).
\newblock {bacondecomp: Goodman-Bacon Decomposition in R}.

\bibitem[\protect\citeauthoryear{Freyaldenhoven, Hansen, and
  Shapiro}{Freyaldenhoven et~al.}{2019}]{freyaldenhoven2019pre}
Freyaldenhoven, S., C.~Hansen, and J.~M. Shapiro (2019).
\newblock Pre-event trends in the panel event-study design.
\newblock {\em American Economic Review\/}~{\em 109\/}(9), 3307--38.

\bibitem[\protect\citeauthoryear{Friedberg}{Friedberg}{1998}]{friedberg1998did}
Friedberg, L. (1998).
\newblock Did unilateral divorce raise divorce rates? evidence from panel data.
\newblock {\em The American Economic Review\/}~{\em 88\/}(3), 608--627.

\bibitem[\protect\citeauthoryear{Gardner}{Gardner}{2021}]{gardnertwo}
Gardner, J. (2021).
\newblock Two-stage differences in differences.
\newblock Working paper.

\bibitem[\protect\citeauthoryear{Gentzkow, Shapiro, and Sinkinson}{Gentzkow
  et~al.}{2011}]{gentzkow2011}
Gentzkow, M., J.~M. Shapiro, and M.~Sinkinson (2011).
\newblock The effect of newspaper entry and exit on electoral politics.
\newblock {\em American Economic Review\/}~{\em 101\/}(7), 2980--3018.

\bibitem[\protect\citeauthoryear{Gobillon and Magnac}{Gobillon and
  Magnac}{2016}]{gobillon2016regional}
Gobillon, L. and T.~Magnac (2016).
\newblock Regional policy evaluation: Interactive fixed effects and synthetic
  controls.
\newblock {\em Review of Economics and Statistics\/}~{\em 98\/}(3), 535--551.

\bibitem[\protect\citeauthoryear{Goodman-Bacon}{Goodman-Bacon}{2021}]{goodman2021difference}
Goodman-Bacon, A. (2021).
\newblock Difference-in-differences with variation in treatment timing.
\newblock {\em Journal of Econometrics\/}~{\em 225}, 254--277.

\bibitem[\protect\citeauthoryear{Goodman-Bacon, Goldring, and
  Nichols}{Goodman-Bacon et~al.}{2019}]{bacondecompStata}
Goodman-Bacon, A., T.~Goldring, and A.~Nichols (2019, July).
\newblock {BACONDECOMP: Stata module to perform a Bacon decomposition of
  difference-in-differences estimation}.

\bibitem[\protect\citeauthoryear{Graham and Powell}{Graham and
  Powell}{2012}]{graham2012identification}
Graham, B.~S. and J.~L. Powell (2012).
\newblock Identification and estimation of average partial effects in
  “irregular” correlated random coefficient panel data models.
\newblock {\em Econometrica\/}~{\em 80\/}(5), 2105--2152.

\bibitem[\protect\citeauthoryear{Imai and Kim}{Imai and
  Kim}{2021}]{imai2021use}
Imai, K. and I.~S. Kim (2021).
\newblock On the use of two-way fixed effects regression models for causal
  inference with panel data.
\newblock {\em Political Analysis\/}~{\em 29\/}(3), 405--415.

\bibitem[\protect\citeauthoryear{Jakiela}{Jakiela}{2021}]{jakiela2021simple}
Jakiela, P. (2021).
\newblock Simple diagnostics for two-way fixed effects.
\newblock arXiv preprint arXiv:2103.13229.

\bibitem[\protect\citeauthoryear{Jord{\`a}}{Jord{\`a}}{2005}]{jorda2005estimation}
Jord{\`a}, {\`O}. (2005).
\newblock Estimation and inference of impulse responses by local projections.
\newblock {\em American economic review\/}~{\em 95\/}(1), 161--182.

\bibitem[\protect\citeauthoryear{Kahn-Lang and Lang}{Kahn-Lang and
  Lang}{2020}]{kahn2020promise}
Kahn-Lang, A. and K.~Lang (2020).
\newblock The promise and pitfalls of differences-in-differences: Reflections
  on 16 and pregnant and other applications.
\newblock {\em Journal of Business \& Economic Statistics\/}~{\em 38\/}(3),
  613--620.

\bibitem[\protect\citeauthoryear{Liu, Wang, and Xu}{Liu
  et~al.}{2021}]{liu2021practical}
Liu, L., Y.~Wang, and Y.~Xu (2021).
\newblock A practical guide to counterfactual estimators for causal inference
  with time-series cross-sectional data.
\newblock arXiv preprint arXiv:2107.00856.

\bibitem[\protect\citeauthoryear{Manski and Pepper}{Manski and
  Pepper}{2018}]{manski2018right}
Manski, C.~F. and J.~V. Pepper (2018).
\newblock How do right-to-carry laws affect crime rates? coping with ambiguity
  using bounded-variation assumptions.
\newblock {\em Review of Economics and Statistics\/}~{\em 100\/}(2), 232--244.

\bibitem[\protect\citeauthoryear{Marcus and Sant'Anna}{Marcus and
  Sant'Anna}{2021}]{Marcus2020}
Marcus, M. and P.~H. Sant'Anna (2021).
\newblock The role of parallel trends in event study settings: An application
  to environmental economics.
\newblock {\em Journal of the Association of Environmental and Resource
  Economists\/}~{\em 8\/}(2), 235--275.

\bibitem[\protect\citeauthoryear{Rambachan and Roth}{Rambachan and
  Roth}{2019}]{rambachan2019honest}
Rambachan, A. and J.~Roth (2019).
\newblock An honest approach to parallel trends.
\newblock Working paper.

\bibitem[\protect\citeauthoryear{Rios-Avila, Sant'Anna, and
  Callaway}{Rios-Avila et~al.}{2021}]{csdidStata}
Rios-Avila, F., P.~Sant'Anna, and B.~Callaway (2021).
\newblock Csdid: Stata module for the estimation of difference-in-difference
  models with multiple time periods.

\bibitem[\protect\citeauthoryear{Roth}{Roth}{2021}]{roth2019pre}
Roth, J. (2021).
\newblock Pre-test with caution: Event-study estimates after testing for
  parallel trends.
\newblock {\em American Economic Review: Insights\/}~{\em forthcoming}.

\bibitem[\protect\citeauthoryear{Roth and Sant'Anna}{Roth and
  Sant'Anna}{2021}]{roth2021efficient}
Roth, J. and P.~H. Sant'Anna (2021).
\newblock Efficient estimation for staggered rollout designs.
\newblock arXiv preprint arXiv:2102.01291.

\bibitem[\protect\citeauthoryear{Roth, Sant'Anna, Bilinski, and Poe}{Roth
  et~al.}{2022}]{Roth2022}
Roth, J., P.~H. Sant'Anna, A.~Bilinski, and J.~Poe (2022).
\newblock What's trending in difference-in-differences? a synthesis of the
  recent econometrics literature.
\newblock arXiv preprint arXiv:2201.01194.

\bibitem[\protect\citeauthoryear{Sant'Anna and Callaway}{Sant'Anna and
  Callaway}{2021}]{didR}
Sant'Anna, P. and B.~Callaway (2021, December).
\newblock did: Treatment effects with multiple periods and groups in r.

\bibitem[\protect\citeauthoryear{Schmidheiny and Siegloch}{Schmidheiny and
  Siegloch}{2020}]{Schmidheiny20}
Schmidheiny, K. and S.~Siegloch (2020).
\newblock On event studies and distributed-lags in two-way fixed effects
  models: Identification, equivalence, and generalization.
\newblock ZEW Discussion Paper 20-01.

\bibitem[\protect\citeauthoryear{S{\l}oczy{\'n}ski}{S{\l}oczy{\'n}ski}{2020}]{sloczynski2020should}
S{\l}oczy{\'n}ski, T. (2020).
\newblock When should we (not) interpret linear iv estimands as late?
\newblock arXiv preprint arXiv:2011.06695.

\bibitem[\protect\citeauthoryear{Stevenson and Wolfers}{Stevenson and
  Wolfers}{2006}]{stevenson2006bargaining}
Stevenson, B. and J.~Wolfers (2006).
\newblock Bargaining in the shadow of the law: Divorce laws and family
  distress.
\newblock {\em The Quarterly Journal of Economics\/}~{\em 121\/}(1), 267--288.

\bibitem[\protect\citeauthoryear{Sun}{Sun}{2020}]{EVENTSTUDYWEIGHTS}
Sun, L. (2020, September).
\newblock {EVENTSTUDYWEIGHTS: Stata module to estimate the implied weights on
  the cohort-specific average treatment effects on the treated (CATTs) (event
  study specifications)}.

\bibitem[\protect\citeauthoryear{Sun}{Sun}{2021}]{eventstudyinteractStata}
Sun, L. (2021, August).
\newblock {EVENTSTUDYINTERACT: Stata module to implement the interaction
  weighted estimator for an event study}.

\bibitem[\protect\citeauthoryear{Sun and Abraham}{Sun and
  Abraham}{2021}]{abraham2018}
Sun, L. and S.~Abraham (2021).
\newblock Estimating dynamic treatment effects in event studies with
  heterogeneous treatment effects.
\newblock {\em Journal of Econometrics\/}~{\em 225}, 175--199.

\bibitem[\protect\citeauthoryear{Wolfers}{Wolfers}{2006a}]{wolfers2006did}
Wolfers, J. (2006a).
\newblock Did unilateral divorce laws raise divorce rates? a reconciliation and
  new results.
\newblock {\em American Economic Review\/}~{\em 96\/}(5), 1802--1820.

\bibitem[\protect\citeauthoryear{Wolfers}{Wolfers}{2006b}]{Wolfersdata}
Wolfers, J. (2006b).
\newblock Replication data for: Did unilateral divorce laws raise divorce
  rates? a reconciliation and new results.
\newblock Technical report, Nashville, TN: American Economic Association
  [publisher], 2006. Ann Arbor, MI: Inter-university Consortium for Political
  and Social Research [distributor], 2019-12-07.

\bibitem[\protect\citeauthoryear{Wooldridge}{Wooldridge}{2021}]{wooldridge2021two}
Wooldridge, J. (2021).
\newblock Two-way fixed effects, the two-way mundlak regression, and
  difference-in-differences estimators.
\newblock Available at SSRN 3906345.

\bibitem[\protect\citeauthoryear{Zhang and de~Chaisemartin}{Zhang and
  de~Chaisemartin}{2020}]{did_multiplegtR}
Zhang, S. and C.~de~Chaisemartin (2020, October).
\newblock {did\_multiplegt: DID Estimation with Multiple Groups and Periods in
  R}.

\bibitem[\protect\citeauthoryear{Zhang and de~Chaisemartin}{Zhang and
  de~Chaisemartin}{2021}]{twowayfeweightsR}
Zhang, S. and C.~de~Chaisemartin (2021, May).
\newblock {TwowayFEWeights: Estimation of the Weights Attached to the Two-Way
  Fixed Effects Regressions in R}.

\end{thebibliography}

\end{document}